\documentclass[10pt,conference]{IEEEtran}
\usepackage{cite}
\usepackage{amsmath,amssymb,amsfonts}
\usepackage{algorithmic}
\usepackage{subfig}
\usepackage{graphicx}
\usepackage{textcomp}
\usepackage{xcolor}
\usepackage[hyphens]{url}
\usepackage{fancyhdr}
\usepackage{hyperref}
\usepackage{multirow}
\usepackage{soul}
\sethlcolor{green}

\graphicspath{{./figs/}}
\usepackage{circledsteps}
\pgfkeys{/csteps/fill color=black}
\pgfkeys{/csteps/inner color=white}

\newcommand{\hpcayear}{2024}

\title{A Lightweight, Compiler-Assisted Register File Cache for GPGPU}

\def\hpcacameraready{} 

\newcommand\hpcaauthors{Mojtaba Abaie Shoushtary$\dagger$, Jose Maria Arnau$\dagger$, Jordi Tubella Murgadas$\dagger$, Antonio Gonzalez$\dagger$}
\newcommand\hpcaaffiliation{Polytechnic University of Catalonia (UPC)\\Barcelona, Spain$\dagger$}
\newcommand\hpcaemail{}

\author{
  \ifdefined\hpcacameraready
    \IEEEauthorblockN{\hpcaauthors{}}
      \IEEEauthorblockA{
        \hpcaaffiliation{} \\
        \hpcaemail{}
      }
  \else
    \IEEEauthorblockN{\normalsize{HPCA \hpcayear{} Submission
      \textbf{\#\hpcasubmissionnumber{}}} \\
      \IEEEauthorblockA{
        Confidential Draft \\
        Do NOT Distribute!!
      }
    }
  \fi 
}

\fancypagestyle{camerareadyfirstpage}{%
  \fancyhead{}
  
  \fancyhead[C]{
    \ifdefined\aeopen
    \parbox[][12mm][t]{13.5cm}{\hpcayear{} IEEE International Symposium on High-Performance Computer Architecture (HPCA)}    
    \else
      \ifdefined\aereviewed
      \parbox[][12mm][t]{13.5cm}{\hpcayear{} IEEE International Symposium on High-Performance Computer Architecture (HPCA)}
      \else
      \ifdefined\aereproduced
      \parbox[][12mm][t]{13.5cm}{\hpcayear{} IEEE International Symposium on High-Performance Computer Architecture (HPCA)}
      \else
      \parbox[][0mm][t]{13.5cm}{\hpcayear{} IEEE International Symposium on High-Performance Computer Architecture (HPCA)}
    \fi 
    \fi 
    \fi 
    \ifdefined\aeopen 
      \includegraphics[width=12mm,height=12mm]{ae-badges/open-research-objects.pdf}
    \fi 
    \ifdefined\aereviewed
      \includegraphics[width=12mm,height=12mm]{ae-badges/research-objects-reviewed.pdf}
    \fi 
    \ifdefined\aereproduced
      \includegraphics[width=12mm,height=12mm]{ae-badges/results-reproduced.pdf}
    \fi
  }
  \fancyfoot[C]{}
}
\begin{document}
\maketitle


  \thispagestyle{plain}
  \pagestyle{plain}
\newcommand{\hpcaheight}{0mm}
\ifdefined\eaopen
\renewcommand{\hpcaheight}{12mm}
\fi


\begin{abstract}

Modern GPUs require  an enormous register file (RF) to store the context of thousands of active threads. It consumes considerable energy and contains multiple large banks to provide enough throughput. Thus, a RF caching mechanism can significantly improve the performance and energy consumption of the GPUs by avoiding reads from the large banks that consume significant energy and may cause port conflicts.

This paper introduces an energy-efficient RF caching mechanism called Malekeh that repurposes an existing component in GPUs' RF to operate as a cache in addition to its original functionality. In this way, Malekeh minimizes the overhead of adding a RF cache to GPUs. Besides, Malekeh leverages an issue scheduling policy that utilizes the reuse distance of the values in the RF cache and is controlled by a dynamic algorithm. The goal is to adapt the issue policy to the runtime program characteristics to maximize the GPU's performance and the hit ratio of the RF cache. The reuse distance is approximated by the compiler using profiling and is used at run time by the proposed caching scheme. We show that Malekeh reduces the number of reads to the RF banks by 46.4\% and the dynamic energy of the RF by 28.3\%. Besides, it improves performance by 6.1\% while adding only 2KB of extra storage per core to the baseline RF of 256KB, which represents a negligible overhead of 0.78\%.

\end{abstract}
\section{Introduction and Motivation}\label{sec:introduction}

Modern GPUs implement fine-grain context switching by relying on a large register file (RF) that stores the context of a very large number of active threads. As the RF is huge and serves many read/write requests, it also contributes to a considerable share of dynamic energy consumption. For instance, NVIDIA V100 architecture has a 20MB RF, representing 56\% of on-chip storage, \cite{V100whitepaper} which consumes about 24\% of the dynamic energy of the chip \cite{accelwattch}.

The RF must also deliver high bandwidth as well as large capacity. Therefore, it consists of enormous single-ported banks to avoid the area and energy overhead of many ports required for high throughput \cite{SinglePortedRF}. Read requests to the same bank are serialized, which affects read bandwidth and latency.

Applications that are more sensitive to operand read bandwidth would be more penalized by bank conflicts. Conventionally, GPUs solve this issue by serving concurrently the operand read requests from different instructions to better utilize the banks' read bandwidth. To implement this, the RF has several Operand Collector Units (OCUs) \cite{OCU}, each of which buffers source operands of one instruction before being dispatched to execution. Each OCU is responsible for fetching and storing the source operands of the allocated instruction.

\begin{figure}[t]
    \centering
    \includegraphics[width=\columnwidth]{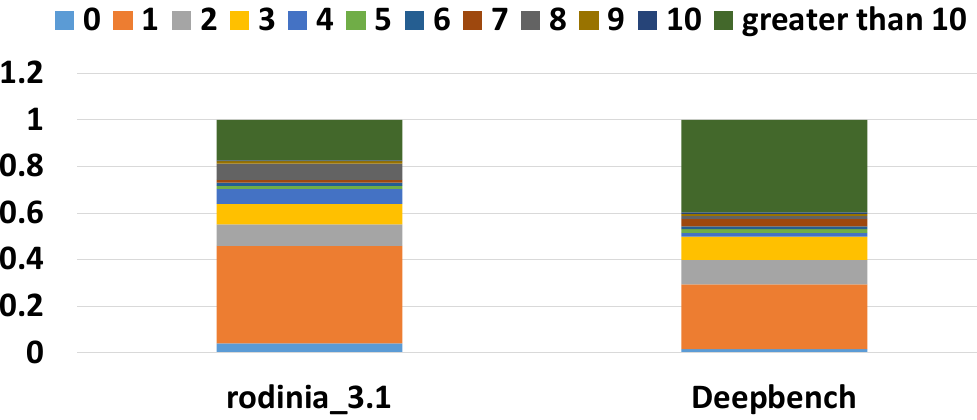}
    \caption{Reuse distance distribution of register values used at least once}
    \vskip -0.25in
    \label{fig:reuse_dist}
\end{figure}

\begin{figure*}[t]
    \centering
    \includegraphics[width=\textwidth,trim = .05cm .05cm .05cm .05cm, clip]{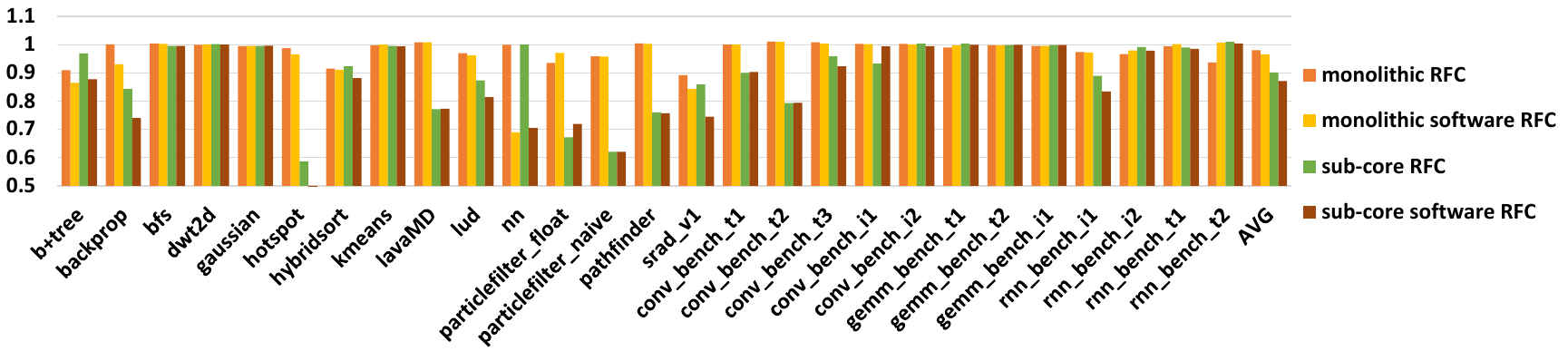}
    \caption{IPC impact of two-level schedulers of RFC and software RFC in monolithic and sub-core based architectures normalized to their baseline}
    \label{fig:two_level}
\end{figure*}

Although OCUs reduce the bank conflict's detrimental effect, many of them are needed to overcome the severe bank conflict issue of modern GPUs. Modern GPUs suffer more from bank conflicts because their Streaming Multiprocessor (SM) is partitioned into sub-cores \cite{SubCore}. Each sub-core has private RF banks, OCUs, issue schedulers, and execution units (EUs) \cite{SubCore, Maxwellwhitepaper,P100whitepaper,V100whitepaper,Turingwhitepaper,A100whitepaper,H100whitepaper,Adawhitepaper}. Partitioning SM into sub-cores reduces the area and energy of the SM \cite{GTX750whitepaper} but increases the chance of bank conflicts \cite{SubCore}. For instance, Volta architecture has only 2 RF banks per sub-core \cite{DissectVolta} which has a high probability of bank conflicts.

A na\"ive solution to tackle severe bank conflicts in sub-cores is to increase the number of OCUs. However, scaling the number of OCUs has a high overhead. The overhead is due to requiring a bigger crossbar, which delivers read operands from banks to the OCUs, and more operand buffering storage of added OCUs. For example, increasing the number of OCUs per sub-core from 2 to 8 improves performance by 7.1\% on average but increases the area and power of the RF by $1.74\times$ and $2.83\times$ respectively \cite{SubCore}. 

A recently proposed issue scheduler tackles the bank conflict problem in sub-cores and avoids the overhead of scaling OCUs \cite{SubCore}. However, it only addresses the bank conflict problem, not the high energy consumption of reading operands from the large RF banks. A possible solution that reduces both the bank conflicts and energy consumption of the RF is using an energy-efficient RF cache, that is small and closer to the Execution Units (EUs), and thus,  more energy efficient than the large RF banks to serve read requests. It also reduces bank conflicts when a request is served directly from the cache since it does not need to be sent to the RF banks.

Previous works, referred to as BOW \cite{BOW}, RFC \cite{RFC}, software RFC \cite{RFC-Compiler}, and LTRF \cite{LTRF}, propose a RF cache for GPUs that unlike modern GPUs do not have sub-cores and do not support Tensor Cores. Sub-cores have been the trend for SM architectures since their appearance in Maxwell architecture \cite{Maxwellwhitepaper,GTX750whitepaper,P100whitepaper,V100whitepaper,TU100whitepaper,A100whitepaper,H100whitepaper,Adawhitepaper}. Tensor cores are also an indispensable accelerator in modern GPUs for General Matrix Multiplication (GEMM) operations \cite{V100whitepaper,TU100whitepaper,A100whitepaper,H100whitepaper,Adawhitepaper}, the building block for modern machine learning and linear algebra computations. These previous proposals, as we will show later in this paper, are rather inefficient or unfeasible in modern sub-core-based architectures with tensor cores.

Malekeh overcomes these limitations in the following way.

First, Malekeh implements the cache inside OCUs instead of adding a separate RF cache, unlike RFC, software RFC, and LTRF. Adding the RF cache apart from OCUs has the overhead of the RF cache itself, routing networks delivering the traffic into/out of the cache, and pipeline modifications to integrate the RF cache. Instead, Malekeh repurposes the OCUs to operate as both RF cache and OCU to avoid such overheads. OCUs already have storage space to buffer source operands of an instruction, but they do not function as a cache. By turning the OCUs into an RF cache, Malekeh reuses the values already stored in the OCU and seamlessly integrates the RF cache into the pipeline.

Second, Malekeh time shares the RF cache storage among warps instead of having a private RF cache per warp, as BOW does. Having a dedicated RF cache per warp makes easy to implement compiler optimizations since the content of the cache is deterministic, as there are no conflicting access from other warps and GPUs issue instructions in-order. However, it has a high area and power overhead. Having a private RF cache, as BOW does, requires as many OCUs as the number of warps. It means that a GPU with 64 warps per SM and four sub-cores, similar to A100 \cite{A100whitepaper,CudaGuide}, requires 16 OCUs per sub-core. Even ignoring the overhead of the added cache storage, increasing the number of OCUs from 2 to 16 has a power overhead of $5.19\times$ \cite{SubCore} in the RF.

Third, Malekeh captures temporal locality through a fully associative cache with a very small number of entries, unlike BOW, which requires many more entries to support modern tensor core instructions. This is because BOW requires a buffer to store all sources and destinations of a few instructions within a sliding window. In that design, the more instructions in the sliding window or the more source and destinations per instruction, the larger the buffer must be. Tensor core instructions have both of these characteristics. They have a high number of source and destination registers and a relatively high reuse distance, which requires scaling the sliding window size significantly to capture locality. In particular, Fig. \ref{fig:reuse_dist} depicts the reuse distance of register values for Deepbench \cite{Deepbench} and Rodinia \cite{Rodinia} compiled to Turing native ISA \cite{SASS}. Deepbench has a high frequency of tensor core instructions, 65.6\% for conv\_bench, and shows higher reuse distances. Fig. \ref{fig:reuse_dist} also shows that more than 40\% of reuses in Deepbench have a distance greater than 10. In addition, in Turing's native ISA, tensor core instructions can have up to eight 4B source and destination registers \cite{TENSOR,TENSOR1,PTX}. Putting this all together, a buffer to store source and destination registers of 11 tensor core instructions for 32 threads in a warp requires $32\times11\times8\times4B=11KB$. This overhead is avoided by Malekeh by using a much smaller number of entries and relying on more efficient cache management policies guided by reuse distances.

Fourth, Malekeh uses a one-level scheduling policy, unlike RFC, software RFC, and LTRF, which use a two-level scheduler to reduce overhead. RF cache schemes using two-level schedulers allocate the RF cache space only to a subset of warps called active set. Only the warps in the active set can issue instructions in each cycle. The performance of these two-level schedulers is highly sensitive to the policy that determines which and how many warps are active. In a modern architecture with sub-cores, the number of warps per sub-core is low, and thus, the number of warps in the active set is even lower and limits the capability to hide short latency stalls. A two-level scheduler causes unnecessary stalls in the issue cycle when a warp is ready yet not in the active set. Fig. \ref{fig:two_level} shows the IPC impact of the two-level schedulers proposed in RFC and software RFC for two different architectures with 32 warps and an active set of 8 warps per SM. In one architecture, monolithic, there is only one scheduler issuing one instruction per cycle common in early GPU architectures such as  Tesla \cite{Tesla}. In another architecture, sub-core, warps are distributed on four sub-cores, each having one scheduler as it is the case in modern architectures such as Turing \cite{TU100whitepaper}. Each scheduler manages only 8 warps, 2 of which are in the active set. The results show a very important drop in performance of 12.9\% for software RFC and 9.9\% for RFC on average in the sub-core architecture. Some applications such as hotspot, suffer a 50\% performance loss using software RFC in the sub-core architecture. Although the monolithic architecture experiences a small performance drop of 2.1\% for RFC and 3.5\% for software RFC on average, it is not negligible for some applications such as nn that suffer a 31.1\% IPC drop using software RFC. In conclusion, previously proposed RF cache schemes that leverage a two-level scheduler incur in a very important performance loss, especially in modern architectures that use sub-cores, which outweighs any potential benefit in RF energy reduction. On the other hand, Malekeh not only does not cause any penalties, but also provides a significant performance improvement.  

To sum up, avoiding the above-mentioned inefficiencies requires Malekeh not to increase the number of OCUs, minimally increase the entries in the RF cache and use a single-level scheduling policy. Dealing with such restrictions is challenging and requires highly effective management policies.

To implement effective RF cache management policies, Malakeh leverages information about the reuse distance provided by the compiler and includes a dynamic scheme that delays the issue of some instructions when it is expected to increase the cache hit ratio but at the same time is not causing a performance penalty. We show that a simple approximation of the reuse distance is enough to exploit the full potential of the Malakeh caching scheme, achieving an average RF cache hit ratio of 46.4\%. We also show that Malakeh single-level scheduler enhanced with the dynamic issue delay scheme is highly effective and not only does not incur any penalty but it provides an average 6.1\% performance improvement.

To the best of our knowledge, Malekeh is the first RF cache scheme designed to address the severe bank conflict and energy consumption of modern GPU architectures equipped with sub-cores and tensor cores. Malekeh targets both general-purpose and modern machine-learning applications which extensively make use of tensor cores. 

In summary, the contributions of this paper are the followings:

\begin{itemize}
    \item Introducing a lightweight RF caching mechanism that leverages the already existing OCUs of modern GPU architectures equipped with sub-cores and support for tensor core instructions.
    \item Proposing the design of novel caching and scheduling policies tailored to maximize energy-saving and performance.
    \item Devising a practical approximation of the reuse distance of each register operand computed by the compiler and exploited by the hardware at run time.
    \item Analyzing the overall energy saving and performance provided by the system. Results show 6.1\% IPC improvement and 28.3\% reduction in the dynamic energy of the RF on average.
\end{itemize}

This paper is organized as follows. Section \ref{sec:baseline} covers the baseline RF architecture in GPUs. Section \ref{sec:architecture}
describes Malekeh's architecture and
its ISA extension. Malekeh policies are explained in section \ref{sec:policies}. Section \ref{sec:methodology} explains the evaluation methodology. Section \ref{sec:evaluation} presents the evaluation results and analysis of the mechanism. The
related work and their comparison to this work are done in
section \ref{sec:related}. This work is concluded in section \ref{sec:conclusions}.

\section{Baseline RF Architecture}\label{sec:baseline}
\begin{figure}[t]
    \centering
    \includegraphics[width=\columnwidth]{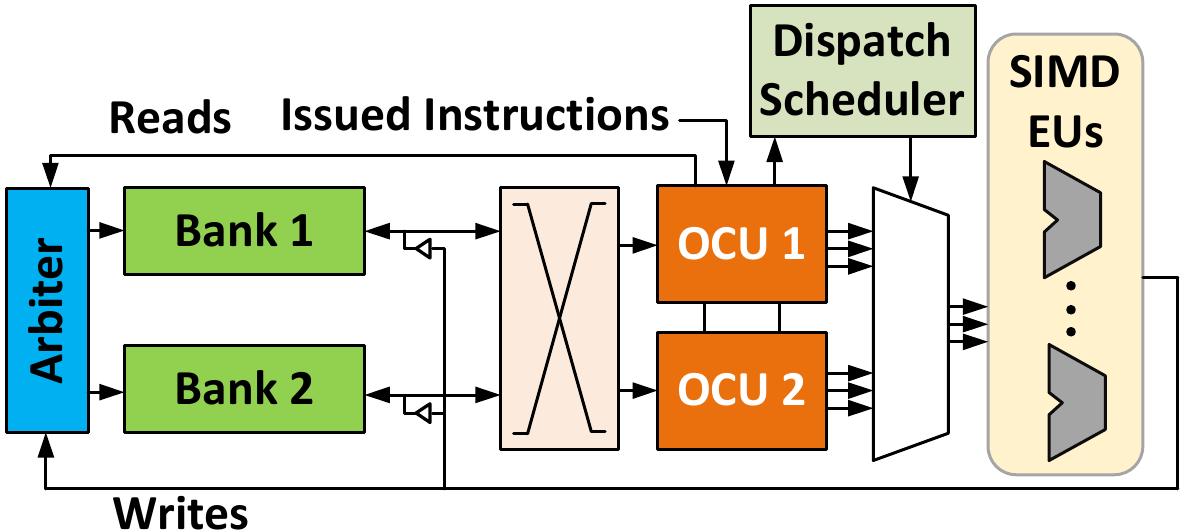}
    \caption{RF microarchitecture in GPUs \cite{gpgpusim3_manual,OCU}}
    \label{fig:RF}
\end{figure}

Our baseline models a Turing SM architecture \cite{Turingwhitepaper} with four sub-cores. Each sub-core has an RF with two banks \cite{DissectVolta}, two OCUs \cite{OCU,SubCore}, an arbiter unit, and a dispatch scheduler, as depicted in Fig. \ref{fig:RF}. The crossbar delivers read register values to OCUs; the arbiter unit resolves the port conflicts, and the dispatch scheduler selects the ready instructions for dispatching to the SIMD EUs.

The banks are single-ported and serve only one read/write request per cycle. Writes always have priority and are granted access immediately but reads requiring the same port of the bank or OCU are conflicting. Conflicting reads are stored in a FIFO queue associated with each bank. Every cycle, the oldest request of each queue is granted access by the arbiter unit only if the port of the OCU and the bank it needs are available.

Each OCU buffers the source operands of an instruction in up to
6 source operand slots to support tensor core instructions \cite{TENSOR,TENSOR1, PTX}. Each source operand slot contains a valid bit, a ready bit, and a data field. The valid bit is set if the source operand slot is used. The
ready bit is set once the fetched data from the banks is
delivered to the OCU and stored in the data field. The data field
is a 128B vector register storing 4B data per thread for a
warp with 32 threads.

The issue scheduling and OCU allocation policies determine which instruction occupies which OCU. A free OCU will randomly be selected by the OCU allocation policy as soon as the issue scheduler selects a new instruction. Once the OCU is allocated to a new instruction, it generates the source operand read requests and pushes them into the RF bank queues.

The instruction in the OCU will be a candidate for dispatch once all of its source operands are fetched from the RF banks, that is when the ready bit is set for all valid source operand slots. Once the dispatch scheduler dispatches the instruction of an OCU, the OCU will be released and reallocated later to another instruction.
\section{Malekeh Architecture and ISA Extenstion} \label{sec:architecture}
\begin{figure}[t]
    \centering
    \includegraphics[width=.99\columnwidth]{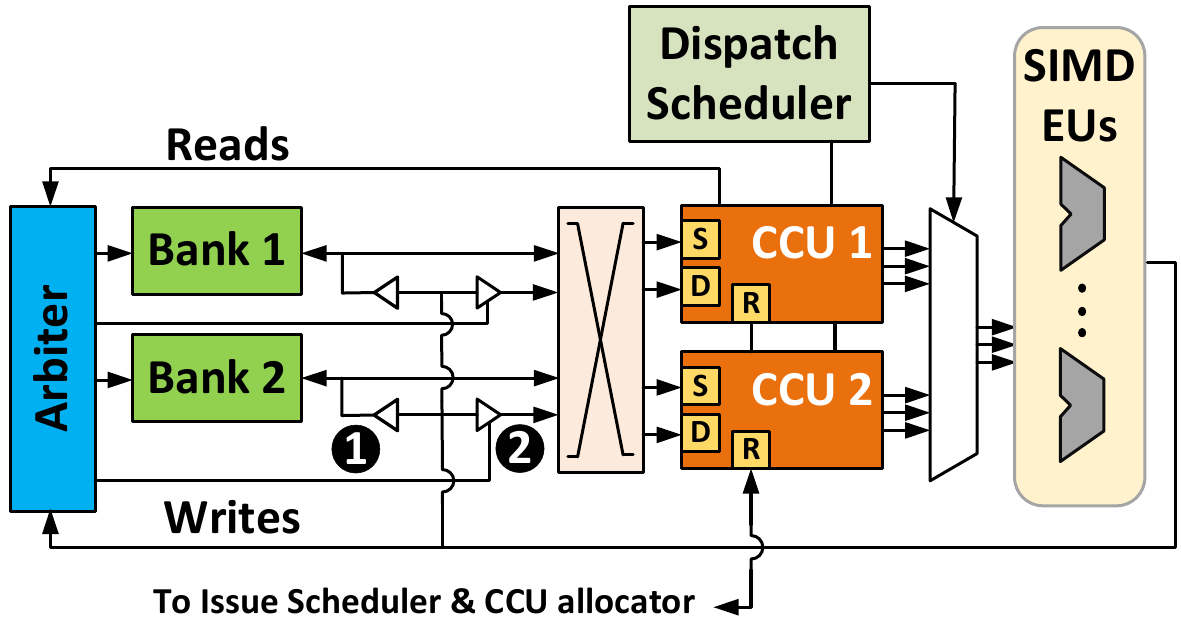}
    \caption{Malekeh RF microarchitecture}
    \label{fig:Malekeh_RF_Microarchitecture}
\end{figure}

Malekeh requires some small hardware support and an ISA extension to pass the register reuse distance information from the compiler to the hardware. Section \ref{sec:reuse_distance} explains the calculation of reuse distance, while section \ref{sec:malekeh_rf_architecture} details the RF hardware architecture modifications. This includes the replacement of OCUs with Caching Collector Units (CCUs), as described in section \ref{sec:malekeh_ccu_architecture}.

\subsection{Computing Reuse Distance}\label{sec:reuse_distance}
The reuse distance is the number of dynamic instructions between a source or destination register and its immediate reuse. It is computed by the compiler, encoded in the instructions, and passed to the hardware. Thus, it may potentially increase the program size. To keep this potential overhead minimal, we propose to use a binary approximation. This binary approximation encodes reuse distances as just two possible values: far and near. The reuse distances above a predefined threshold, RTHLD, are designated as far, and those lower than the threshold are considered as near. Therefore, the reuse distance used in Malekeh is only one bit.

Although RTHLD can be different for kernels, applications, policies, and even during runtime, we used only one RTHLD during the whole execution and empirically found 12 provides the best results for our benchmarks. 

The exact reuse distance of each operand is unknown at compile time in general due to a twofold reason. First, the reuse distance of operands reused across multiple basic blocks depends on the actual control flow at run time. Second, For operands reused within a basic block, the reuse distance is not deterministic because modern GPUs support interleaved execution \cite{V100whitepaper} which may interleave both paths of a branch during run time. A reuse in  a basic block may be spaced by instructions of another independent path.

To handle the above issue, the compiler collects profiling statistics for the reuse of each operand in a kernel regarding how many times its reuse is far and how many times is near. Then, it marks each operand's reuse as the most common one encountered during profiling. Profiling is offline for the first few warps of each kernel. We verified that profiling only a few warps (around 0.01\%) produces accurate results, very close to profiling the whole execution, so Malekeh adopted this low-overhead partial profiling.

\subsection{Malekeh RF Microarchitecture}\label{sec:malekeh_rf_architecture}
The RF architecture in Malekeh has three main differences compared to the baseline, as depicted in Fig. \ref{fig:Malekeh_RF_Microarchitecture}.

First, the OCUs are replaced by CCUs. A CCU is a unit that retrieves and stores all source operands of an instruction, like an OCU, and includes extensions to reuse them.
 
Second, CCUs operands can be updated with the results of some instructions. This increases the reuse possibilities, but it may have significant energy and area overhead if done for all the results. This would require multiple write ports and a significantly bigger crossbar since multiple instructions of the same warp may simultaneously reach the write-back stage. We empirically verified that adding one single write-back port in each CCU provides almost the same benefit as an unbounded number of ports. To reduce energy consumption, Malekeh uses a write filtering policy to avoid writes that are unlikely to be reused. To filter writes, a new tri-state buffer \Circled{\textbf{2}} beside buffer \Circled{\textbf{1}} is added, which is controlled by the RF arbiter. The arbiter squashes some specific write requests based on their reuse distance.

Third, information about the CCUs status is provided to the issue scheduler and the CCU allocator to improve their effectiveness, as described in detail in section \ref{sec:Issue_Scheduling}.

\subsection{CCU Microarchitecture}\label{sec:malekeh_ccu_architecture}
\begin{figure}[t]
    \centering
    \includegraphics[width=.90\columnwidth]{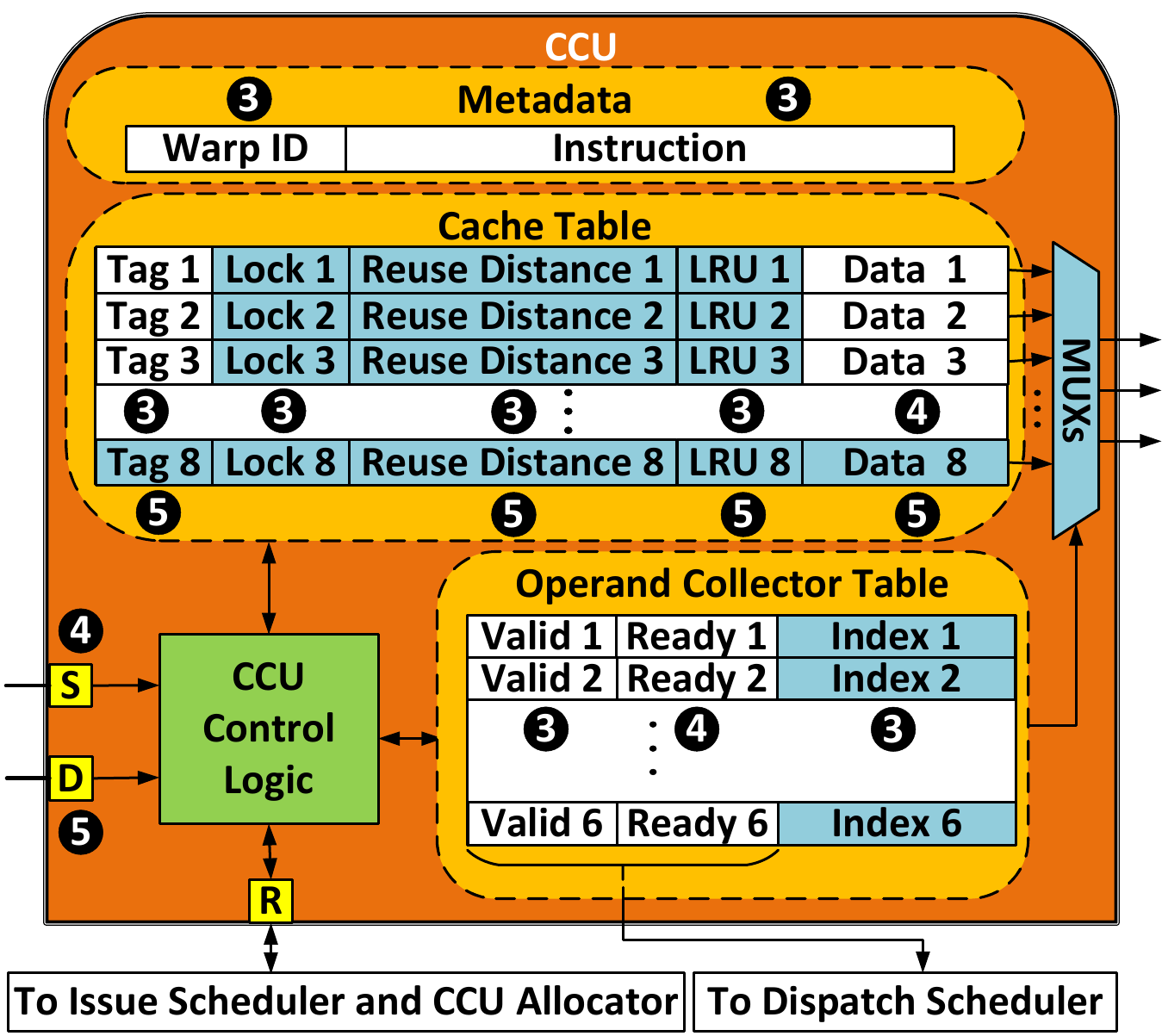}
    \caption{CCU microarchitecture}
    \label{fig:Malekeh_CC_Microarchitecture}
\end{figure}

To exploit the temporal locality in the RF accesses, conventional OCUs are augmented with caching capabilities. The baseline OCUs already have storage to keep registers’ data and some metadata. Malekeh turns the OCU
into a cache while keeping its previous functionality by adding
extra fields and modifying the control logic. This new unit is called CCU, and its microarchitecture is shown in Fig. \ref{fig:Malekeh_CC_Microarchitecture}. The components in white color were already in the OCU; the rest are the new hardware required by Malekeh.

Each CCU has three ports named by letters: a) port S for receiving the source operand values from the RF banks; b) port D to receive the value written to the instruction's destination; and c) port R to communicate information between the CCU and the Issue scheduler and CCU allocator, i.e., Warp ID and the reuse distance of the live values in the CCU.

In addition, each CCU contains five main parts that are described below.

\noindent\textbf{Metadata:} The metadata contains information required to dispatch the instruction occupying the CCU and information needed in the following stages of the pipeline such as the Warp ID, target EU, etc. This metadata is the same as in the baseline architecture.

\noindent\textbf{Cache Table(CT):} The CT is the component that provides the caching capability. The higher the number of CT entries,  the greater its capability to reuse but also its overhead. Its cost is proportional to the number of entries, whereas its extra reuse potential increases in a sub-linear manner, and beyond a given size, it reaches a point of diminishing returns. Our analysis shows that eight entries is the sweet spot of this curve, and it is what we consider in this work. Note that the baseline OCU has six entries so Malakeh simply adds two additional ones.

The CT contains the data values and some information required to manage them as a cache: tags, lock bits, reuse distances, and Least Recently Used (LRU) priority information. 

Data fields store 128B register values, representing the largest area in each CCU. The tag is the register identifier. In CUDA \cite{CUDA} the maximum number of addressable registers per thread is 256; therefore, the tag is only one byte. 

The lock bit indicates that the register is allocated to one of the source operands of the instruction occupying the CCU, pending to be dispatched. It informs the replacement policy to avoid replacing this register since all the source operand values are required when the instruction in the CCU is dispatched to the corresponding EU.

The reuse distance is computed by the compiler as explained in section \ref{sec:reuse_distance} and shows how far the next reuse is. Information about the reuse distance of the values in each CCU and the warp id is made available to the issue scheduler and CCU allocator through port R. In particular, a single bit is sent indicating whether the CCU contains any near value in the CT or not. 

The LRU field represents the priority order needed for an LRU cache replacement policy. In our case, three bits are enough to show the order of replacement of eight blocks.

\noindent\textbf{Operand Collector Table (OCT):}
The OCT has 6 slots (to support tensor core instructions), one per source operand, to keep track of the status of the source operands of the instruction occupying the CCU. Each source operand slot has a valid and ready bit with the same functionality as the baseline, besides an index field. The index field stores a pointer to the CT entry containing the source operand's data, thus implementing indirect indexing. This indirect indexing eliminates the need for redundant data values and improves the effective use of the available small cache space. Since, in our design, the CT is of size 8, the index field requires 3 bits.

\noindent\textbf{MUXs:} Once an instruction has all its source operands ready and is selected to be dispatched to the corresponding execution unit, these MUXs are used to deliver its source operands to the input latches of the execution unit. The size of these MUXs is proportional to the number of entries in the CT. Therefore, keeping the CT size small reduces the MUXs overhead.

\noindent\textbf{CCU Control Logic:}  It is responsible for controlling the CCU
to support the four operations detailed in section \ref{sec:CCU_Operations} below.

\subsubsection{CCU Operations}\label{sec:CCU_Operations}
Four operations are required in the CCU: 1) CCU allocation; 2) receiving a source operand value; 3) receiving a destination operand write request, and 4) dispatching the instruction occupying the CCU to the corresponding EU.

\noindent\textbf{CCU allocation:} When a CCU is allocated to a new instruction, the fields denoted by \Circled{\textbf{3}} will be used and updated as follows. 

First, the CT is flushed if the warp id of the new instruction is different from the warp id of the previous one, stored in metadata. This is necessary as warps have a private register set.

Second, the metadata part will be updated with the information of the new instruction received over port R.

Third, for all the source operands of the instruction, a tag check is performed to see whether they are already present in the CCU. In case of a miss, a new CT entry will be allocated for the operand, by replacing one of the entries according to the replacement policy. 

Fourth, Certain CT fields require updating, including setting the lock bit, updating the reuse distance with the new instruction's reuse distances, and updating the LRU fields. While theoretically, the reuse distance field of all registers should be decreased for each new instruction, we propose to only update the reuse distance of the registers belonging to the new instruction. Our analysis confirms that this simplification is effective and avoids the overhead of updating the reuse distance of all the registers in the CT for every CCU allocation. 

Fifth, read requests to the RF banks for all the missed values are sent.

Sixth, the OCT fields are updated. The valid bit is set, and the index field is updated to point to the corresponding CT entry for each source operand. The ready bit is set for the values found in the CT. For the remaining operands, the ready bit will be set once the value is received over port S.

\noindent\textbf{Receiving a source operand value:} As soon as a requested source operand value arrives over port S, its corresponding ready bit in the OCT is set, and the data is stored in the data value field. The involved fields and ports are denoted by \Circled{\textbf{4}}.

\noindent\textbf{Receiving a destination write request:} The ports and the fields used in this operation are marked with \Circled{\textbf{5}}. The register id is looked up in the CT to check whether the register is present. In case of a miss, an entry is replaced and allocated to this write. The data is copied and the reuse distance and LRU fields are updated. Unlike source operands, there is no need to set the lock bit, since the cache block can be replaced with no harm. To avoid cache pollution and use the small space of the cache efficiently, the cache block allocation for destinations is postponed to the cycle when the instruction reaches the write-back stage of the pipeline, unlike source operands that are allocated when the instruction is issued to the CCU.

\noindent\textbf{Dispatching an instruction to an execution unit: } An instruction is ready to be dispatched as soon as all its source operands have been received, that is when all the valid source operands have the ready bit set.
\section{Malekeh Policies}\label{sec:policies}
Malekeh uses efficient and lightweight management policies. A key aspect of these policies is  that they leverage the reuse distance information provided by the compiler.

Even though our baseline is limited to two OCUs, Malekeh supports systems with any number of OCUs. Our analysis demonstrates their efficacy as the number of OCUs increases. This section presents the policies for any number of OCUs.

Section \ref{sec:Cache_Policices} explains the RF cache policies, and section \ref{sec:Issue_Scheduling} presents the issue scheduling/CCU allocation policy used by Malekeh. 
\subsection{RF Cache Policies}\label{sec:Cache_Policices}
Efficient RF cache management policies are crucial for maximizing Malekhe's hit ratio and performance, given its small cache size of 8 blocks per CCU, and the need to support tensor core instructions. Tensor core instructions require up to 6 sources and 2 destinations and tend to have long reuse distances, which makes the design of effective policies more challenging. The cache policies should avoid cache pollution and suboptimal replacements to minimize unnecessary capacity misses. In addition, they must minimize issue stalls due to having all CCUs occupied. Section \ref{sec:replacement_policy} explains Malekeh's replacement policy, and section \ref{sec:write_policy} elaborates on the write policy. 
\subsubsection{Replacement Policy}\label{sec:replacement_policy}

Malekeh's cache replacement policy gives priority to keeping registers with near reuse distance.

The policy first excludes all registers with the lock bit set, as they are source operands needed when the instruction is dispatched to the execution units. Among the remaining ones, it randomly selects one among those registers with a far reuse distance, if any exists. If there are no far registers, the replacement is chosen according to the LRU policy. This policy reduces the sub-optimal replacement of values with near reuse by LRU while maintaining its benefits. 

\subsubsection{Write Policy}\label{sec:write_policy}

In Malekeh, when a warp instruction completes execution and reaches the write-back stage, the destination register is updated in the RF banks if its data is not present in any CCU. This can occur if the CCU that originally contained the data was reallocated to another warp while the instruction was still in the EU pipelines, causing the data to be flushed. Otherwise, the data is also written in the CCU only if its reuse is near. Writes with far reuse distance are not cached to reduce cache pollution and energy waste caused by writes that may be replaced without being used.

Multiple writes to the same CCU may arrive simultaneously since multiple instructions of the same warp with different latencies may reach the write-back stage simultaneously, even if they started execution in different cycles. To avoid the overhead of multiple write ports, Malekeh selects the one of the writes with near reuse distance, if any. We empirically verified that a single port provides benefits very close to an unbounded number of ports.

Note that since the RF banks are always updated for all the write requests, any CCU's cache can be flushed at any time with no additional action required.

\subsection{Issue Scheduling and CCU Allocation Policy}\label{sec:Issue_Scheduling}

The scheduling policy consists of two parts: one that selects a warp based on priority and another that uses a CCU allocation policy to choose the target CCU. Each scheduler issues an instruction only if there is a ready warp and a free CCU; otherwise, the issue stage is stalled. Moreover, Malakeh uses a lightweight waiting mechanism that stalls the issue sometimes, to enhance the effectiveness of the CCU allocation policy and maximize both IPC and hit ratio.

Fig. \ref{fig:Issue_Scheduler} depicts these policies which are described in detailes in the next sections. In particular,  section \ref{sec:warp_priorities} elaborates on Malekhe's warp priorities, section \ref{sec:ccu_allocation_policy} details the CCU allocation policy, and section \ref{sec:dynamic_algorithm} explains a dynamic algorithm to determine the waiting time.

\subsubsection{Warp Priorities}\label{sec:warp_priorities}

The Greedy Then Oldest (GTO) warp priority scheme has been proven to effectively reduce contending memory accesses to the memory hierarchy in GPUs \cite{CCWS}. GTO prioritizes the warp that issued the most recent instruction and if it does not have a ready instruction, then it selects the oldest warp with a ready instruction. However, this scheme is agnostic to the state of the RF cache and results in a low RF hit ratio.

Malekeh proposes a new warp priority denoted by \Circled{\textbf{1}} in Fig. \ref{fig:Issue_Scheduler}. Similar to GTO, the highest priority is to the same warp if it is ready. The remaining warps are then divided into two categories: those having data in CCUs and the rest. Within each category, they are prioritized by their age. This approach retains the benefits of GTO by prioritizing the last issued warp overall and respecting the warps' ages in each category while it increases RF cache reuse by prioritizing warps with data in the CCUs.
\subsubsection{CCU Allocation Policy}\label{sec:ccu_allocation_policy}
\begin{figure}[t]
    \centering
    \includegraphics[width=.99\columnwidth]{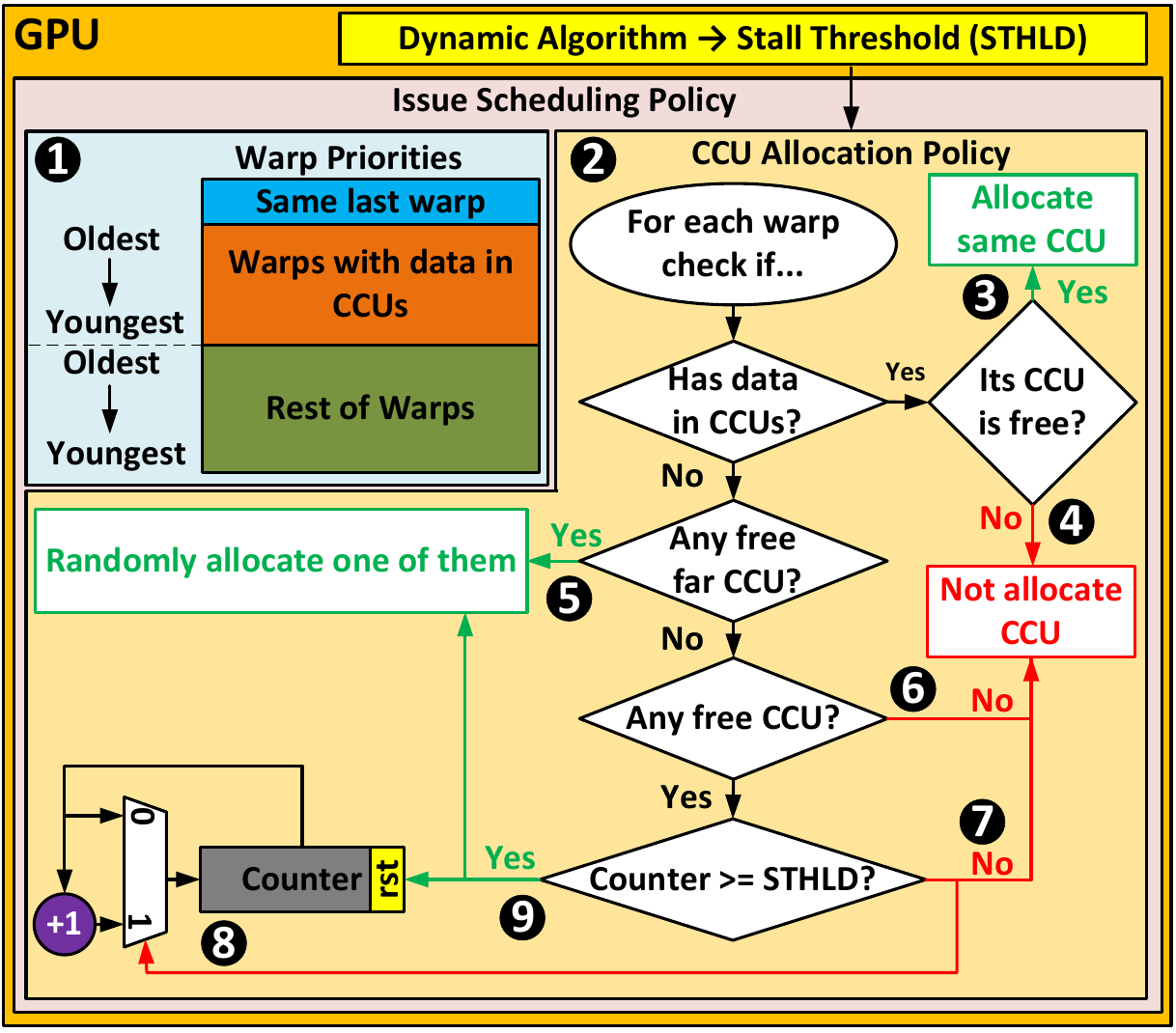}
    \caption{Malekeh issue scheduling policy definition and its interaction with the dynamic algorithm}
    \label{fig:Issue_Scheduler}
\end{figure}
Malekeh CCU allocation policy is in box \Circled{\textbf{2}} in Fig. \ref{fig:Issue_Scheduler}  . In this graph, cases when a CCU is allocated are highlighted in green, while other cases, where no CCU is allocated, are highlighted in red.

Malekeh allocates the same CCU to a warp that has data in a CCU only if the CCU is free \Circled{\textbf{3}}. If the CCU is already occupied, no other CCU will be allocated \Circled{\textbf{4}}. By using this approach, Malekeh can avoid the overhead of implementing a complex cache coherence protocol, as no warp can have data in more than one CCU.

When none of the CCUs is allocated to the warp, Malekeh assigns a target CCU based on the reuse distance of values in the CCUs and flushes the cache. This flush reduces the potential hit ratio if the reuse distance of the flushed data is near, so Malekeh randomly selects a free CCU that contains only values having far reuse (i.e. far CCU), if any exists \Circled{\textbf{5}}. If all CCUs are occupied \Circled{\textbf{6}}, no allocation is made. If some CCUs are free but they contain some values with near reuse, allocating any of them to the new warp would potentially harm the hit ratio but not making any allocation may harm performance. To deal with this trade-off, Malekeh employs a waiting mechanism that uses a per-core counter and a per-GPU threshold, STHLD. If the counter is lower than STHLD, no CCU is allocated \Circled{\textbf{7}} and the counter is increased \Circled{\textbf{8}}. Otherwise, a randomly selected free CCU is allocated, and the counter will be reset \Circled{\textbf{9}}.

This waiting mechanism postpones the CCU allocation of CCUs that contain near values. During this time, an instruction of an "old" warp having data in any of these CCUs may finish execution and reach the write-back stage, which may resolve a data dependence. After that, the warp currently using this CCU becomes ready and can issue instructions that may reuse the data in CCUs.

Delaying the allocation of CCUs may cause a performance drop only if it lengthens the critical path of the application's execution. However, Malekeh does not cause any performance penalty; in fact, it improves performance (as discussed in section \ref{sec:performance}) since the RF cache reduces bank conflicts.

\subsubsection{Dynamic Algorithm to set STHDL}\label{sec:dynamic_algorithm}

\begin{figure}[t]
    \centering
    \hspace*{\fill}%
    \subfloat[]{\includegraphics[width=0.5\columnwidth]{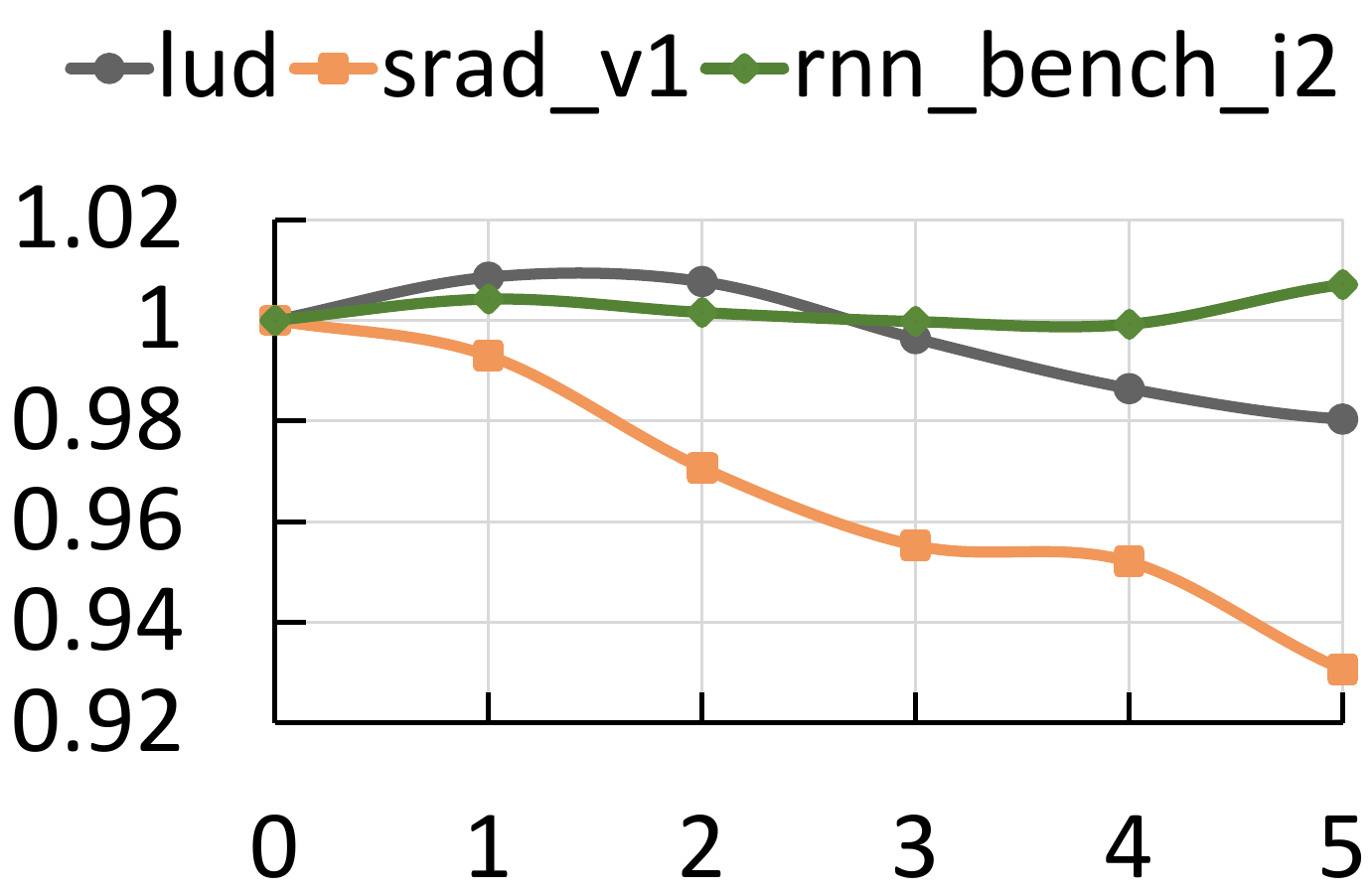}\label{IPC_dynamic}}\hfill
    \subfloat[]{\includegraphics[width=0.5\columnwidth]{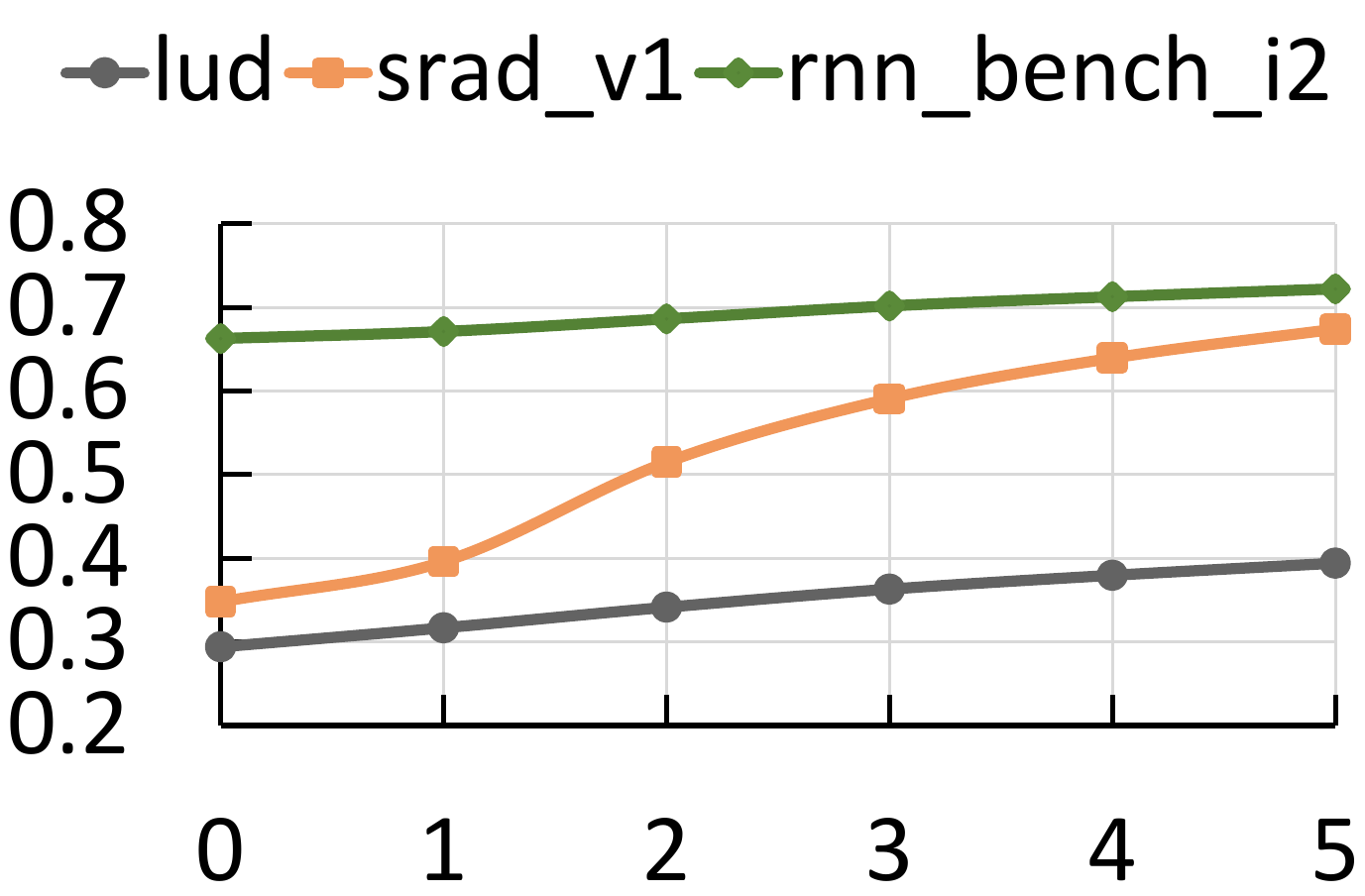}\label{hit_ratio_dynamic}}
    \hspace*{\fill}%
    \caption{IPC and hit ratio of benchmarks ran for 10000 cycles for different STHLD values: \protect\subref{IPC_dynamic} Normalized IPC, \protect\subref{hit_ratio_dynamic} Hit ratio}%
    \label{fig:dynamic}
\end{figure}

Fig. \ref{fig:dynamic} depicts the IPC and hit ratio of three applications when changing STHLD. Although all applications gain hit ratio from higher STHLD values, some sensitive applications, such as srad\_v1 start to lose performance even with small STHLD values, such as STHLD=1. The optimal STHLD is different for each  application, and even for each different phase of the same application. Using  a wrong STHLD may significantly penalize IPC or hit ratio or both. Because of this, Malekeh uses a dynamic algorithm to set the STHLD value. 

The higher STHLD is, the higher the chances of benefiting from reuse, so if we plot the curve of RF cache hit ratio versus STHDL, we will get a monotonic growing curve. On the other hand, higher STHLD values lead to more stalls generated by the CCU allocator, which may damage performance. We can expect that if we increase STHLD we will initially have small fluctuations, until a point when IPC starts dropping dramatically. This point is different for each different code and we will refer to it as the knee point. The region to the left will be called the flat region and the region to the right will de denoted as the steep region. Therefore, the optimal STHLD is at the knee point, since it provides the highest IPC and hit ratio.

Most GPU applications are quite regular or have regular phases, therefore, the characteristics of the application remain similar for long intervals until the application phase changes. Motivated by that, Malekeh’s scheme partitions the execution into equal intervals and set the STHLD at the end of each interval. It uses the IPC measured in the previous and current intervals to modify the STHLD for
the next interval. After several intervals, the STHLD converges to its optimum point.

\begin{figure}[t]
    \centering
    \includegraphics[width=.99\columnwidth]{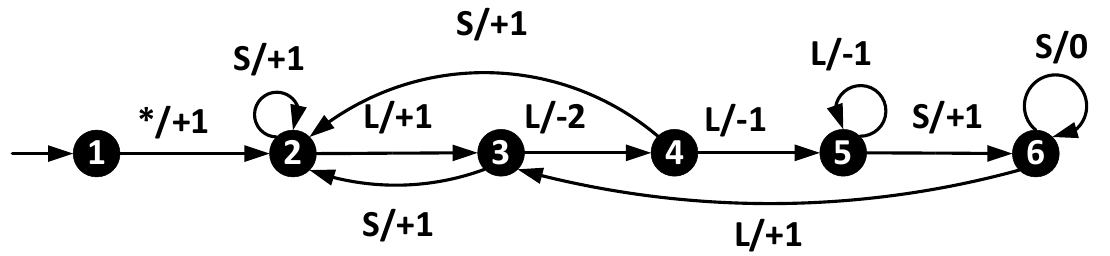}
    \caption{Dynamic algorithm to set STHLD}
    \label{fig:State_Machine}
\end{figure}

\begin{figure}[t]
    \centering
    \hspace*{\fill}%
    \subfloat[]{\includegraphics[width=0.5\columnwidth]{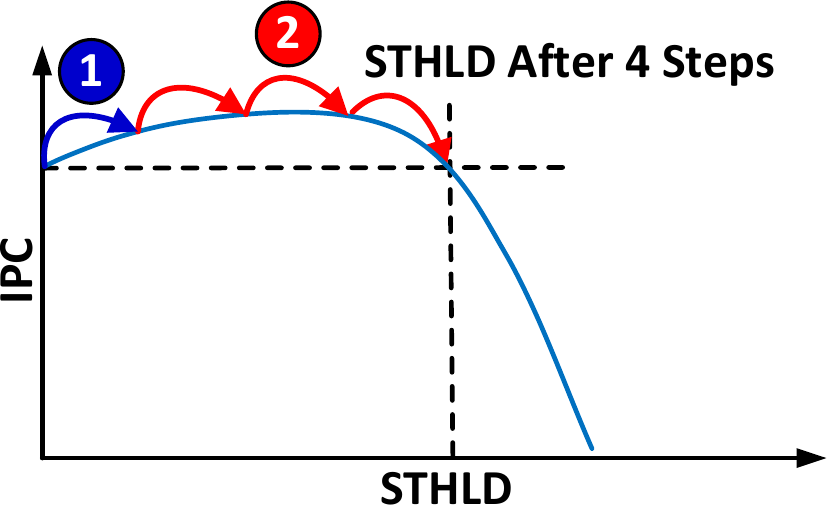}\label{original}}\hfill
    \subfloat[]{\includegraphics[width=0.5\columnwidth]{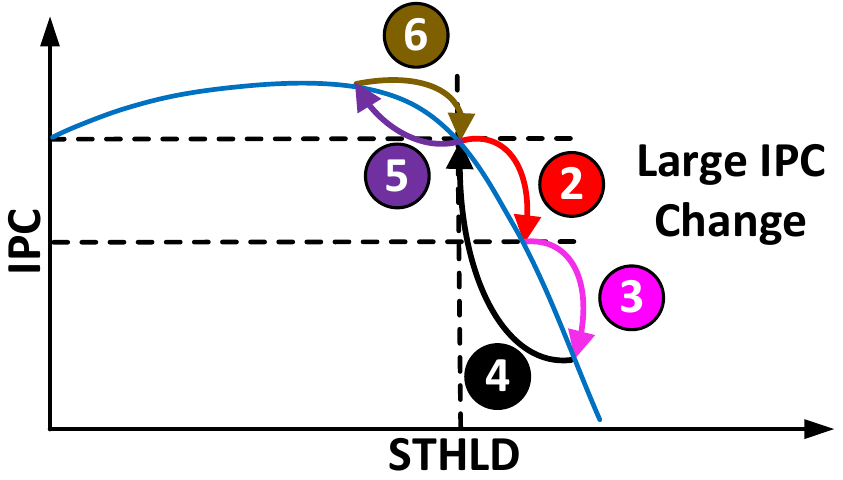}\label{case1}}
    \hspace*{\fill}\\
    \hspace*{\fill}%
    \subfloat[]{\includegraphics[width=0.5\columnwidth]{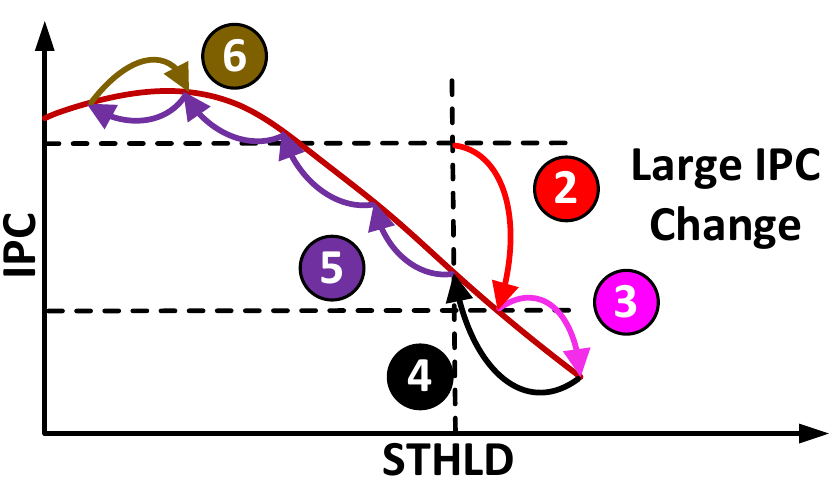}\label{case2}}\hfill
    \subfloat[]{\includegraphics[width=0.5\columnwidth]{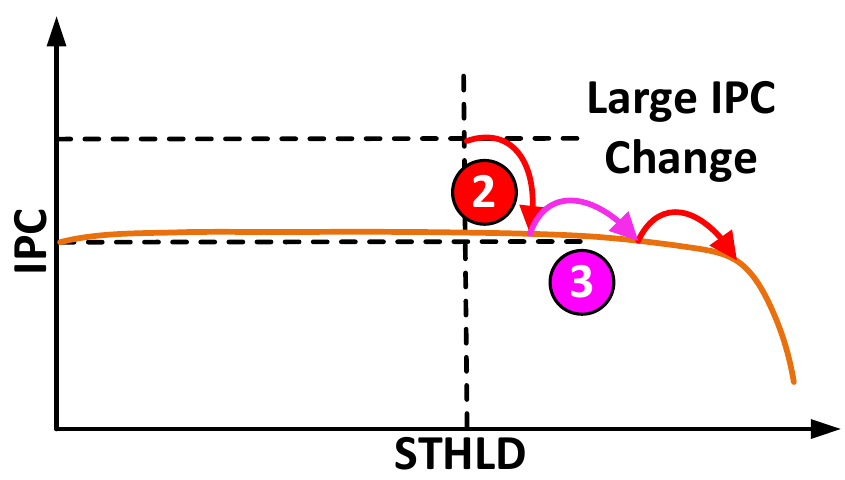}\label{case3}}
    \hspace*{\fill}\\
    \caption{Example of setting STHLD: \protect\subref{original} Initial curve and first four steps, \protect\subref{case1} Next steps when staying in the same curve,\protect\subref{case2} Next steps when transitioning to a curve with a narrower flat region, \protect\subref{case3} Next steps when transitioning to a curve with a wider flat region.}%
    \label{fig:dynamic_sample}
\end{figure}

The dynamic algorithm adopted by Malekeh is described in Fig. \ref{fig:State_Machine} by a finite state machine. It has 6 states and the transitions among the states happen based on the relative difference in IPC of the current and the previous interval. A relative difference of less than 0.02 is considered small and is denoted by S. A relative difference of more than 0.02 is considered large and denoted by L. The value of 0.02 has been empirically set to provide a good performance and hit ratio trade-off. The asterisk symbol shows the transitions happening regardless of the relative IPC difference. Once a transition happens, the STHLD will be increased or decreased by a delta shown on the transition edges.

The behavior of the dynamic algorithm is illustrated in Fig. \ref{fig:dynamic_sample}. In this figure, the corresponding state is shown by a number, and the arrows showing steps taken in each state are colored the same as the circled number showing the corresponding state.
Fig. \ref{fig:dynamic_sample} shows that the dynamic algorithm is designed to walk on the curve and, based on the IPC fluctuation, modify the STHLD until it converges near the knee point. The knee point is the optimum STHLD.

In this example, the algorithm started on the curve shown in Fig. \ref{original}, and after four intervals with a small change in IPC, a large change is detected. The large change could be due to moving to the steep region of the same curve, Fig. \ref{case1}, or a change in the application phase that changes the curve to the curves shown in Fig. \ref{case2} or \ref{case3} for the following next intervals. 

The curve changes as the phase of the application changes because the characteristics of the application differ. The new curve may have a narrower (Fig. \ref{case2}) or wider (Fig. \ref{case3}) flat region. In case of moving to the curve with a narrower flat region, the current STHLD will be in a steep region and the adaptive algorithm will reduce it to reach the STHLD corresponding to the knee point of the IPC curve. On the other hand, in the case of moving to the curve with a long flat region, the current STHLD will be in the flat region, so the dynamic algorithm will increase the STHLD to approach the knee point of the new IPC curve.

In case of a large change, the dynamic algorithm takes a speculative move by increasing STHLD to gain more hit ratio and moves to state 3. If the new phase corresponds to the curve shown in Fig. \ref{case3}, the speculative move was correct and we benefit from the increased IPC and hit ratio due to a higher STHLD in that interval. On the other hand, if the phase has the same curve, Fig. \ref{case1}, or the curve shown in Fig. \ref{case2}, we lose performance since the speculative move was in the steep region where performance decreases when STHLD is increased, but only for one interval. After realizing that this speculative step is detrimental, the scheme decreases STHLD and after some additional steps, it converges to the optimal point. 

The dynamic algorithm remains in state 6 when finding the knee point until a large IPC change is detected.

We empirically found that an interval size of 10000 cycles provides a good trade-off between performance and hit ratio.

\section{Methodology}\label{sec:methodology}

We use Accel-sim \cite{Accelsim} to model a GPU configuration based on the Geforce RTX 2060 \cite{Turingwhitepaper} GPU with the parameters shown in Table \ref{tab:configs}. We scaled down the number of SMs, the size of L2, and the number of memory channels by one-third to compensate for the fact that most of the available benchmarks are much shorter than real applications running on GPUs and to prove the dynamic algorithm's effectiveness.
\begin{table}[h]
  \centering
  \small
  \begin{tabular}{|l|l|}
    \hline
    \textbf{\#SMs} & 10\\
    \hline
    \textbf{\#Threads/Warps per SM } & 1024 / 32\\
    \hline
    \textbf{\#sub-cores per SM } & 4\\
    \hline
    \textbf{RF size per SM} & 256KB\\
    \hline
    \textbf{\#Issue Schedulers per SM} & 4\\
    \hline
    \textbf{Issue Scheduling Policy} & GTO\\
    \hline
    \textbf{L2 size} & 1MB\\
    \hline
    \textbf{L1/Shared Memory per SM} & 64KB\\
    \hline
  \end{tabular}
  \caption{Baseline GPU configuration used in this work}
  \label{tab:configs}
\end{table}

The benchmarks are from Rodinia \cite{Rodinia} and Deepbench \cite{Deepbench} and are listed in Table \ref{tab:benchmarks}. The former suite is representative of general-purpose computing applications, whereas the latter consists of modern deep learning workloads. For Deepbench, various tensor dimensions are used for training and inference. This is shown in the charts by an underscore followed by t, for training, or i, for inference, followed by an id.

\begin{table}[t]
\small
\centering
\begin{tabular}{|l|l|}
\hline
\multicolumn{1}{|l|}{\textbf{Rodinia}\cite{Rodinia}} & \multicolumn{1}{l|}{\textbf{Deepbench}\cite{Deepbench}} \\ \hline\hline
B+tree                                & \multirow{2}{*}{conv\_bench training}  \\
Backprop                              &                                        \\
BFS                                   & \multirow{2}{*}{conv\_bench inference} \\
DWT2D                                 &                                        \\
Gaussian                              & \multirow{2}{*}{gemm\_bench training}  \\
Hotspot                               &                                        \\
Kmeans                                & \multirow{2}{*}{gemm\_bench inference} \\
LavaMD                                &                                        \\
lud                                   & \multirow{2}{*}{rnn\_bench training}   \\
nn                                    &                                        \\
particlefilter float                  & \multirow{2}{*}{rnn\_bench inference}  \\
particlefilter naive                  &                                        \\
pathfinder                            &                                        \\
srad v1                               & \\\hline
\end{tabular}%
\caption{Used benchmarks}
\label{tab:benchmarks}
\end{table}

We used accel-sim in trace mode and annotated the traces with precise reuse distances. Then we provided the simulator with the binary reuse distance to be used by Malekeh.

To evaluate the dynamic energy of the RF, we extended the power model provided in Accelwattch\cite{accelwattch} to model the CCUs. The model includes the arbiter, crossbar, RF banks, and CCUs. For the baseline model, instead of CCUs, the conventional OCUs were modeled.
\section{Evaluation}\label{sec:evaluation}
\subsection{Comparison with RFC, Software RFC, and LTRF}
\begin{figure}[t]
    \centering
    \includegraphics[width=\columnwidth]{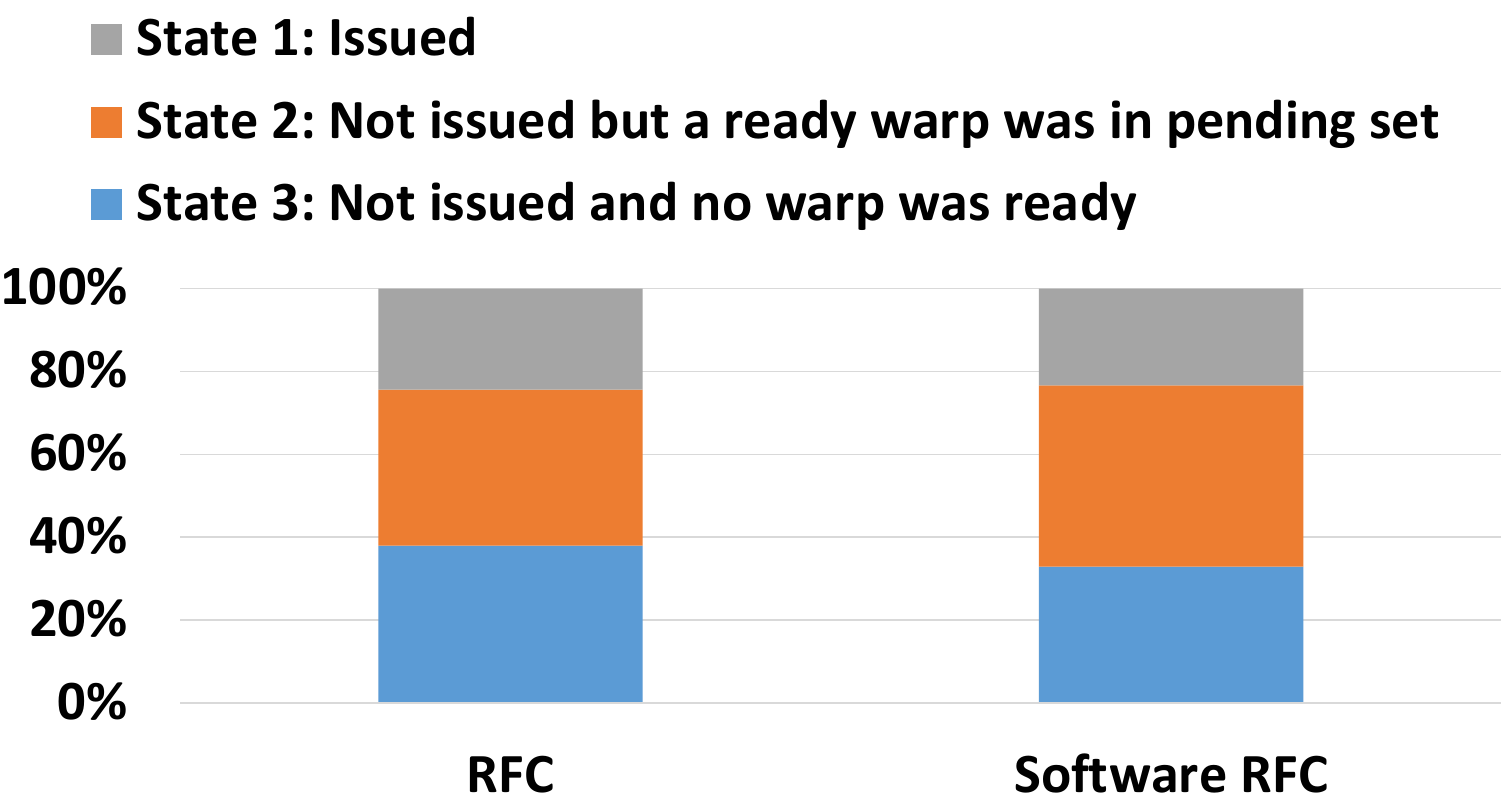}
    \caption{Distribution of the state of RFC and software RFC two-level schedulers in each cycle}
    \label{fig:issue_slot}
\end{figure}

RFC, software RFC, and LTRF use a two-level scheduler to keep the RF cache overhead reasonable. The scheduler divides warps into two sets: active and pending. Only active warps can issue instructions while pending warps need to become active before issuing instructions.

\begin{figure*}[t]
\centering
\begin{minipage}{.25\textwidth}
  \centering
  \includegraphics[width=\linewidth]{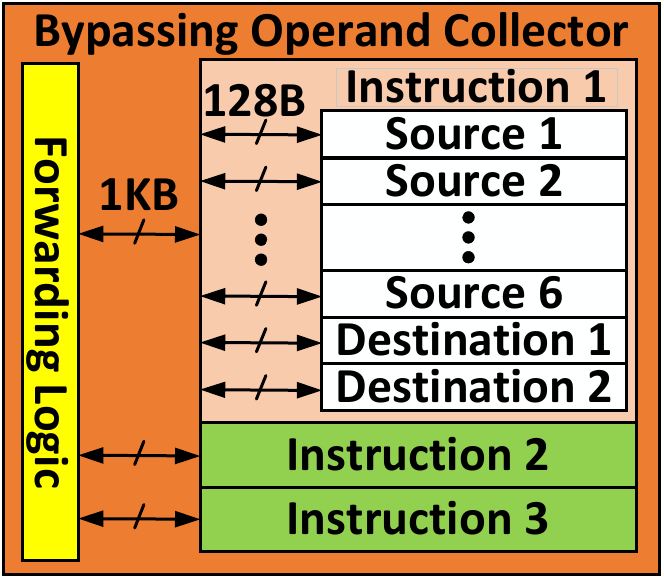}
  \caption{Bypassing Operand Collector (BOC) Architecture}
  \label{fig:boc}
\end{minipage}\hfill
\begin{minipage}{.74\textwidth}
  \centering
  \includegraphics[width=\linewidth]{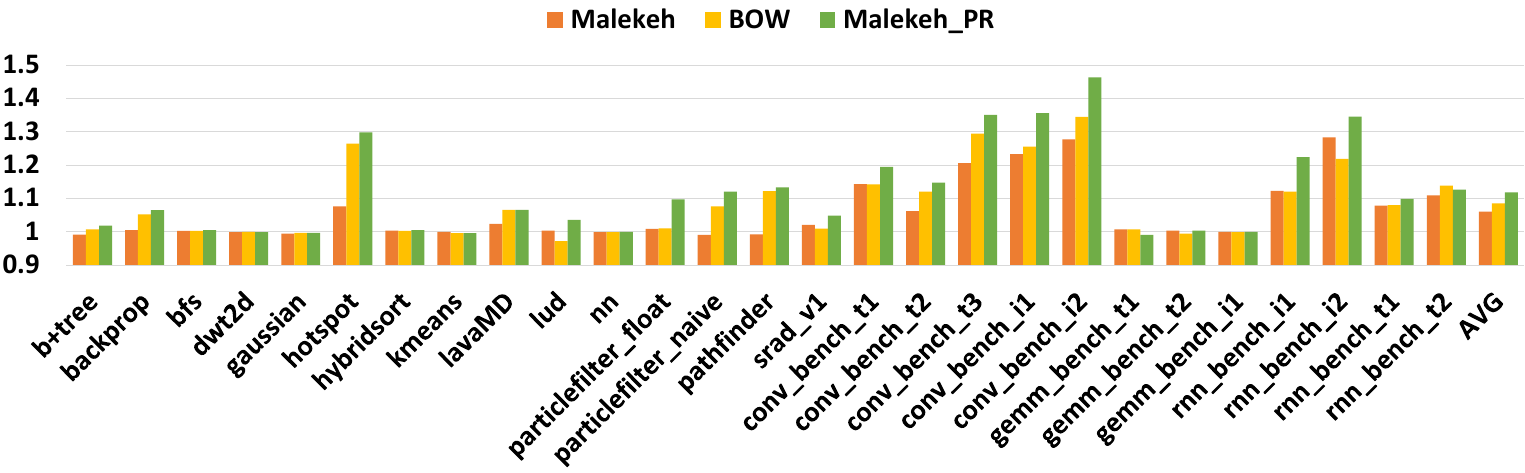}
  \caption{IPC normalized to the baseline}
  \label{fig:IPC}
\end{minipage}
\end{figure*}
\begin{figure*}[t]
    \centering
    \includegraphics[width=\textwidth]{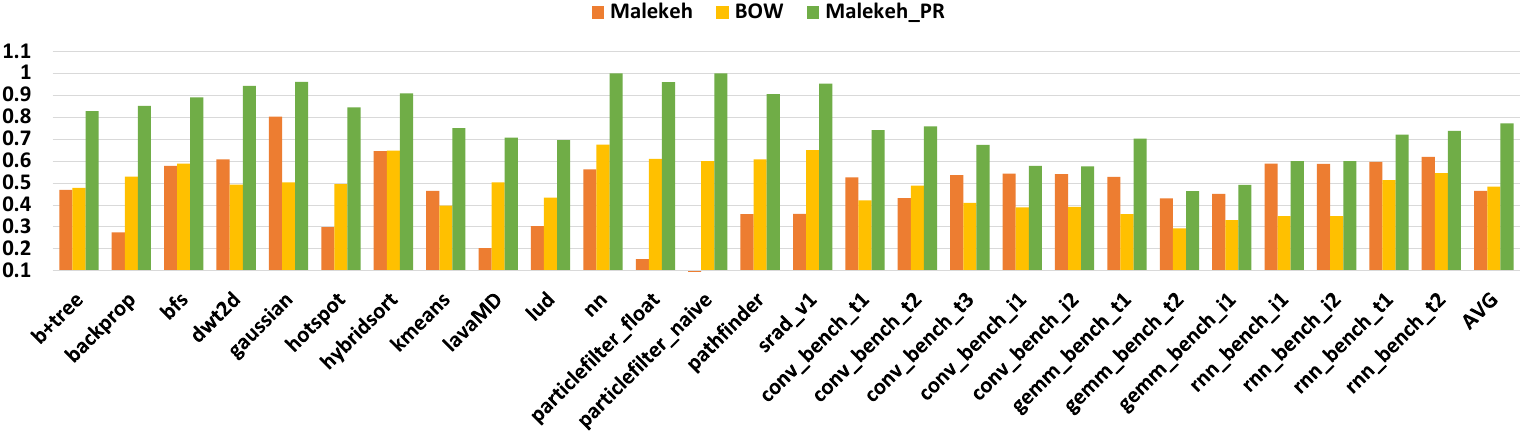}
    \caption{RF cache hit ratio}
    \label{fig:hit_rate}
\end{figure*}
\begin{figure*}[t]
    \centering
    \includegraphics[width=\textwidth]{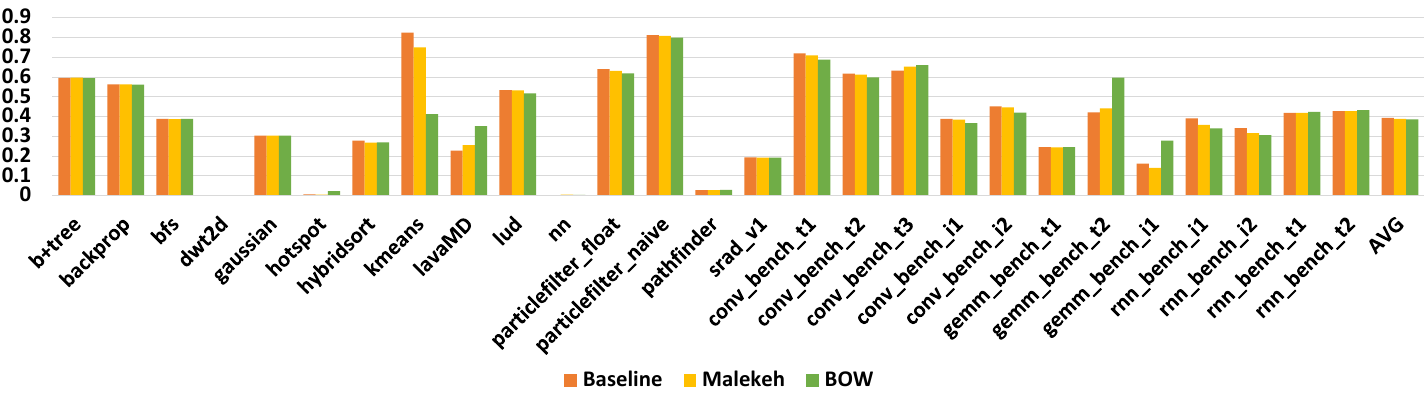}
    \caption{L1 data cache hit ratio}
    \label{fig:L1D_hit_rate}
\end{figure*}
\begin{figure*}[t]
    \centering
    \includegraphics[width=\textwidth]{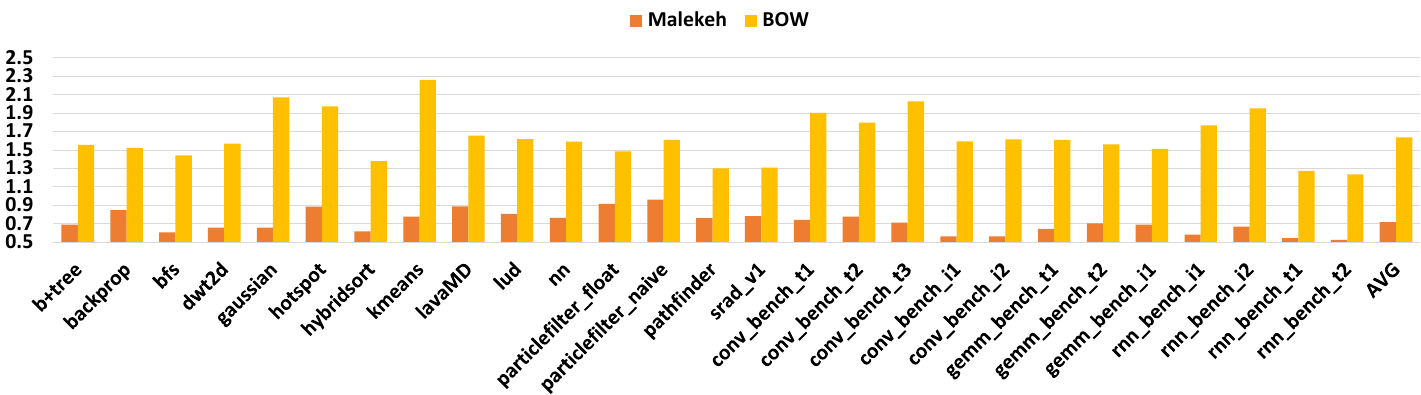}
    \caption{RF dynamic energy normalized to the baseline}
    \label{fig:energy}
\end{figure*}
\begin{figure*}[t]
    \centering
    \includegraphics[width=\textwidth]{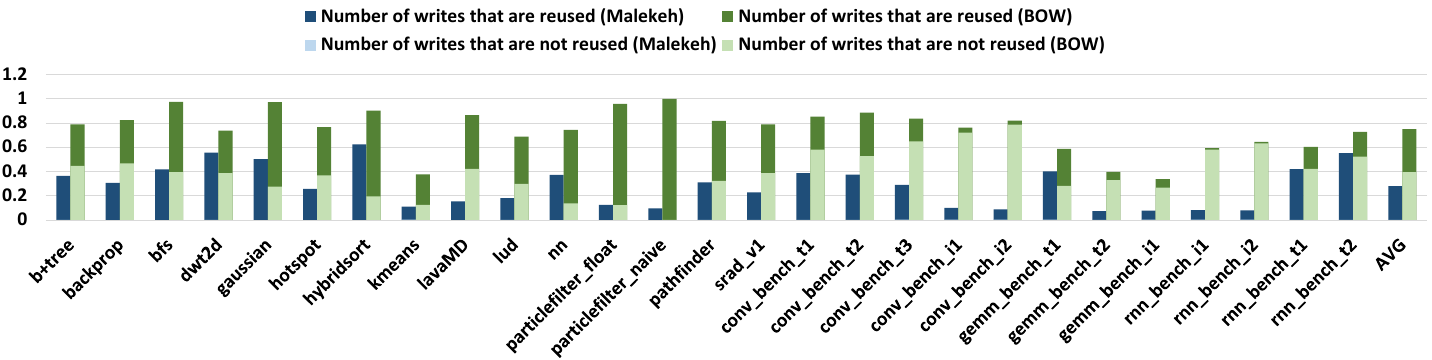}
    \caption{Writes to the RF cache normalized to all writes to the RF}
    \label{fig:writes}
\end{figure*}

\subsubsection{\textbf{RF Dynamic Energy}} 

A two-level scheduler can be in one of three states in each cycle: 1) it issues an instruction, 2) it does not issue an instruction but there is a ready warp in the pending set, or 3) it does not issue an instruction and there are no ready warps. The issue stalls in the second state are avoided by a one-level scheduler and cause a performance penalty.

This penalty is significant when using a two-level scheduler in a modern sub-core-based architecture because the scheduler manages a few warps in a sub-core, and even fewer are active. When very few warps are active, they cannot hide short latencies, and the two-level scheduling policies fail to move warps to the active set soon enough, resulting in the scheduler being in state 2 very often.

Fig. \ref{fig:issue_slot} depicts the average distribution of these states for the two-level schedulers proposed in RFC and software RFC. This experiment was conducted for all benchmarks running on an architecture with the baseline configuration, except that the two-level schedulers were implemented. Both schedulers had six pending and two active warps per sub-core (24 pending and 8 actives per SM). RFC and software RFC are in state 2 for 37.6\% and 43.8\% of the cycles respectively. This leads to an IPC loss of 9.9\% for RFC and 12.9\% for software RFC on average. The IPC drop can be very high in some applications, reaching 41.3\% in RFC and 50.9\% in software RFC for hotspot (Fig. \ref{fig:two_level}). This important IPC loss outweighs the potential benefit of their RF cache scheme for a modern sub-core-based architecture. 

Moreover, in software RFC and LTRF, the compiler is responsible for statically allocating the registers of the cache to values. To correctly perform this allocation, the compiler must know in which order the instructions of each thread will be executed. This was possible in early GPU architectures but it is not in recent architectures that interleave the execution of divergent paths of branches.  This interleaving is decided at runtime and thus it is unknown to the compiler, which prevents it from performing this static allocation.


\subsection{Comparison with BOW}\label{comp_bow}
The BOW scheme is the closest scheme to Malekeh. It captures reuses within a sliding window by replacing OCUs with Bypassing Operand Collectors (BOCs), Fig. \ref{fig:boc}. BOCs buffer sources and destinations of instructions within a sliding window and forward operand values found in the buffer to the next instruction, rather than requesting them from the RF banks.

To support tensor core instructions, a BOC has to buffer 6 sources and 2 destinations per instruction, amounting to $3\times8\times128B=3KB$, for a sliding window of size 3. BOW uses private BOCs per warp to avoid the penalty of time sharing a BOC by multiple warps. This requires a $32\times3=96KB$ BOC buffer, 4 crossbars of $2\times8$ 1024 bits for an SM having 32 warps, and 4 sub-cores (our baseline). Time sharing a BOC among multiple warps increase the reuse distance in a nondeterministic way, as each warp has a private register set, and requires a bigger buffer in BOC to capture temporal locality.

To make a fair comparison, we also evaluate a version of Malekeh having private CCUs per warp, Melekeh\_PR. Detailed comparisons can be found below.

\subsubsection{\textbf{Performance}}\label{sec:performance}
Malekeh improves performance due to the high hit ratio and reducing bank conflicts but on the other hand, it may negatively impact performance when postponing CCU allocation. Fig. \ref{fig:IPC} shows that Malekeh has a negligible IPC loss of 0.8\% in the worst case (b+tree) and for most applications sustains or significantly increases performance. It improves IPC by 6.1\% on average and 28.4\% at maximum for rnn\_bench\_i2.

Compared to Malekeh, BOW offers a higher IPC of 2.43\% on average and 18.8\% in the best case (hotspot). However, achieving this requires a private BOC per warp, which needs 8 BOCs per sub-core amounting to $8\times3KB=24KB$ which is $12\times$ the cache space of Malekeh, and a much bigger crossbar. To be fair, we should compare BOW with Malekeh\_PR, which also has private CCUs per warp. In Fig. \ref{fig:IPC} we can see that Malekeh\_PR provides an IPC 3.3\% higher than BOW on average and 12.7\% in the best case (rnn\_bench\_i2). It only provides a lower IPC for gemm\_bench\_t1 (1.7\%) and rnn\_bench\_t2 (1.2\%). Malekeh\_PR outperforms BOW while requiring much lower cache storage, 33\% of BOW's, and a smaller crossbar for writes.

Although a higher RF hit ratio leads to better performance for most applications, this is not always the case. For instance, particlefilter\_float has a higher RF hit ratio (Fig. \ref{fig:hit_rate}) in BOW yet provides almost the same IPC as Malekeh (Fig. \ref{fig:IPC}). This happens when there is a bottleneck in other stages of the pipeline which limits the IPC gain. In this benchmark, the memory pipeline is the bottleneck and restricts the IPC gains from BOW.

Another case is lud, lud shows a 13\% higher RF cache hit ratio for BOW (Fig. \ref{fig:hit_rate}) but its performance is slightly worse (Fig. \ref{fig:IPC}). This is because BOW and Malekeh have different numbers of BOCs and CCUs that are managed by different scheduling and cache policies which results in different nondeterministic runtime behaviors. This runtime behavior determines the performance. In this case, our analysis shows that the bottleneck is in the memory pipeline and lud does not benefit from a higher hit ratio in the RF cache in BOW (Fig. \ref{fig:hit_rate}). However, Malakeh benefits from a higher hit ratio in the L1 data cache which is 2\% higher than for BOW (Fig. \ref{fig:L1D_hit_rate}).

\subsubsection{\textbf{RF Cache Hit Ratio}} 
The hit ratio of the RF cache depends on its management policies. BOW proposes a private cache per warp to remove interleaving access from other warps, but this comes with significant overhead. BOW manages the cache as a sliding window, which requires a large buffer in each BOC to reuse far values.

In modern architectures supporting tensor cores, register value reuse distances are long and nondeterministic due to the tensor cores API and control flow management policies. Fig. \ref{fig:reuse_dist} shows that 36\% of reuses in Rodinia and 50.2\% in Deepbench have reuse distances farther than 3, requiring a BOC buffer storing more than 3 instructions to exploit these reuses, which would incur in a significant overhead.

Malekeh leverages the runtime reuse distance provided by the compiler to guide the cache management policies and manages to exploit many reuses with a very tiny cache. Its scheduling policy is aware of the reuse distance of values in the cache and uses a dynamic scheme to postpone unnecessary cache flushes to maximize hit ratio and performance. 

Fig. \ref{fig:hit_rate} shows that, on average, Malekeh achieves a hit ratio that is 1.9\% lower than BOW, but with only 2 CCUs per sub-core, which requires to 2KB storage, 8.3\% that of BOW (i.e., $12\times$ times smaller). Malekeh has a higher hit ratio than BOW for most applications, especially Deepbench applications that extensively use tensor cores. However, for some applications like Particle\_filter\_naive, it has a 53.5\% lower hit ratio than BOW due to too many cache flushes triggered by the dynamic scheme to sustain performance. For a more fair comparison, we should compare BOW with Malekeh\_PR, which also has private caches like BOW. Malekeh\_PR significantly improves BOW's hit ratio for all applications, on average by 28.9\% and up to 45.8\% for gaussian, and this is achieved by requiring much smaller cache storage (33\% that of BOW). This greater effectiveness is due to the inability of BOW to capture far reuses as it is managed as a sliding window.

\subsubsection{\textbf{RF Dynamic Energy}} 
Malekeh reduces RF dynamic energy by an average of 28.3\% and a maximum of 47.3\% for rnn\_bench\_t2 (Fig. \ref{fig:energy}). These energy savings are achieved by servicing reads directly from the cache instead of RF banks. In contrast, BOW has higher energy consumption than Malekeh, 92.1\% higher on average and up to $1.48\times$ for kmeans. In fact, BOW has higher energy consumption than the baseline due to its larger crossbar which consumes more energy when delivering source operands from the RF banks to the BOCs. BOCs consume energy even when a source operand is present in the cache, as they need to forward it to the next instruction, and its energy is not negligible due to the large cache size. BOW writes all values in both the RF and the cache, while Malekeh avoids writing into the cache for values with far reuse, which are likely to be replaced before their reuse, and will just pollute the cache and waste energy. Although BOW has a write-back version guided by the compiler, it is not feasible to implement it in modern architectures supporting interleaved execution of divergent paths, since the BOC content is not deterministic and the compiler cannot statically identify which writes should be performed in the cache and which ones in the RF banks.
\begin{figure*}[t]
    \centering
    \includegraphics[width=\textwidth]{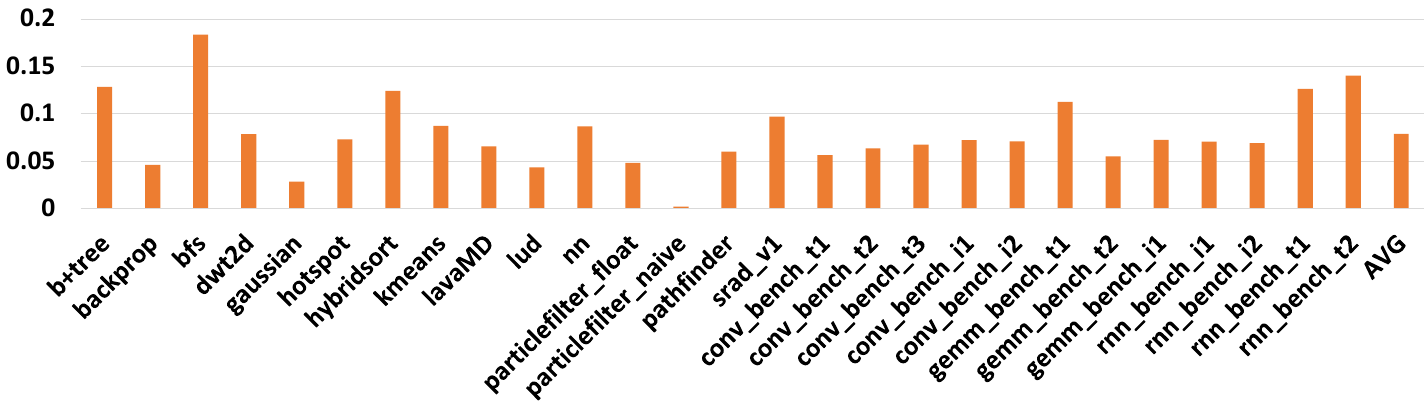}
    \caption{RF cache hit ratio for a system similar to Malekeh with traditional scheduling (GTO) and replacement (LRU) policies}
    \label{fig:naive_hit_rate}
\end{figure*}

Fig. \ref{fig:writes} shows the number of writes to the RF cache normalized to the total writes in the RF. For BOW, we assume that if the allocated slot in the BOC for the destination register slides out of the sliding window before writeback, the produced data will be written only in the RF. This figure confirms that Malekeh has a much smaller number of writes to the cache, and almost all of them put data that is reused. This results in less energy consumption compared to BOW, which pollutes the cache by often writing data in the RF cache that is not reused.

\subsection{Comparison with Traditional Policies}
To show the effectiveness of the designed policies compared to the traditional ones, we measure the hit ratio of the RF cache in a system similar to Malekeh which is governed by traditional scheduling (GTO) and cache replacement (LRU) policies. Fig. \ref{fig:naive_hit_rate} shows this would result in a very poor hit ratio, 7.9\% on average and 18.4\% at maximum for this system. This is due to two main reasons: excessive flushes of the cache when GTO schedules a new warp and frequent replacement of data with near reuse by LRU.

\subsection{Hardware Overhead}
Malekeh is designed to require minimal hardware overhead. It does not increase the number of OCUs, to avoid an increase in the crossbar size in the RF. Malekeh only requires adding two data entries per OCU which amount to 4x2x2x128 = 2KB per SM. This is only 0.78\% of the RF size in the GPU which is 256KB. The other control bits used by CCU represent a negligible overhead.

The adaptive algorithm is also designed to have minimal overhead. It just stores the IPC of the last interval per GPU. Therefore, only one extra register is needed to store the value provided by the performance counter that measures IPC. The logic for the adaptive algorithm is a finite state machine that runs every 10000 cycles, its energy overhead is negligible.

\section{Related Work}\label{sec:related}
\textbf{Register File Caching:} Although using RF cache and hierarchical RF designs are extensively studied for CPUs \cite{CPU1,CPU2,CPU3,CPU4,CPU5,CPU6}, there are few RF cache proposals for GPUs, which have been deeply discussed in sections \ref{sec:introduction} and \ref{sec:evaluation}. 

\noindent\textbf{Issue Scheduling Policy:} 
Apart from the two-level schedulers used in RF cache schemes \cite{LTRF,RFC,RFC-Compiler}, other proposed scheduling policies for GPUs are not designed to improve the effectiveness of an RF cache system. They are designed for different purposes such as reducing bank conflicts \cite{SubCore}, reducing the contention in the L1D cache \cite{CCWS}, tackling branch divergence \cite{LargeWarp}, reducing power consumption \cite{warped-gate}, etc. \cite{OWL,OSPG, DAWS,PATS,CAWS,SAW,ELF,CAWA,BAWS,ORIGAMI,CIAO,CAPAS,GRGS,IGRUT,IPAWS,OAWS,PAWS,PRO,QAWS,SAWS,VWS,BOWS}.

\textbf{RF power and energy saving:}
There are many other research works aimed at reducing the energy and power consumption of RF in GPUs that are not based on caching mechanisms. These include proposals such as using a small Operand Staging Unit backed by global memory instead of the power-hungry RF \cite{Regless}, introducing a compression scheme for the registers \cite{Warped-Compression}, using a rename table to reallocate physical registers with dead values to other warps and reduce RF size \cite{RF-Virtualization}, providing fault mitigation methods for good reliability in the RF while aggressively reducing the supply voltage \cite{Reliabilty-Combating}, and using a compiler-assisted coalescing scheme to reduce the number of read requests to the RF banks \cite{Corf}. Additionally, RF architectures with registers that can switch between on, off, or drowsy states have been proposed \cite{warped-powergating}. Some other proposals leverage high-density and low-power memory cells to build the RF of GPUs \cite{dwm,STT-RAM-sched,edram,RT,STT-RAM-atof,STT-RAM-Gushu,STT-RAM1,RT1}.
\section{Conclusions}\label{sec:conclusions}
In this paper, we have introduced Malekeh, a novel architecture for the RF of sub-core-based GPU architectures that support tensor cores and interleaved execution of divergent branch paths. It leverages the already existing storage in operand collector units to use it as an RF cache and uses compiler information about the reuse distance of register values as a hint in its management policies. As the reuse distance of register values varies during runtime and is nondeterministic, we introduced a lightweight method based on partial profiling to compute it. Besides, a novel instruction scheduling has been proposed that postpones the issue in some beneficial cases based on the reuse distance information. This postponing mechanism is controlled by an adaptive scheme that takes into account the characteristics of the code being executed to maximize both performance and hit ratio. We have shown that Malekeh provides 6.1\% IPC improvement, 46.4\% RF cache hit ratio, and 28.3\% dynamic RF energy saving on average by introducing less than 0.78\% area overhead.


\bibliographystyle{IEEEtranS}
\bibliography{refs}

\begin{thebibliography}{10}
\providecommand{\url}[1]{#1}
\csname url@samestyle\endcsname
\providecommand{\newblock}{\relax}
\providecommand{\bibinfo}[2]{#2}
\providecommand{\BIBentrySTDinterwordspacing}{\spaceskip=0pt\relax}
\providecommand{\BIBentryALTinterwordstretchfactor}{4}
\providecommand{\BIBentryALTinterwordspacing}{\spaceskip=\fontdimen2\font plus
\BIBentryALTinterwordstretchfactor\fontdimen3\font minus \fontdimen4\font\relax}
\providecommand{\BIBforeignlanguage}[2]{{%
\expandafter\ifx\csname l@#1\endcsname\relax
\typeout{** WARNING: IEEEtranS.bst: No hyphenation pattern has been}%
\typeout{** loaded for the language `#1'. Using the pattern for}%
\typeout{** the default language instead.}%
\else
\language=\csname l@#1\endcsname
\fi
#2}}
\providecommand{\BIBdecl}{\relax}
\BIBdecl

\bibitem{gpgpusim3_manual}
T.~M. Aamodt, W.~W. Fung, and T.~H. Hetherington, ``Gpgpu-sim 3.x manual,'' \url{http://gpgpu-sim.org/manual/index.php/Main_Page}, 2022.

\bibitem{warped-powergating}
M.~Abdel-Majeed and M.~Annavaram, ``Warped register file: A power efficient register file for gpgpus,'' in \emph{Proceedings of the 19th IEEE International Symposium on High Performance Computer Architecture (HPCA)}, 2013.

\bibitem{warped-gate}
M.~Abdel-Majeed, D.~Wong, and M.~Annavaram, ``Warped gates: Gating aware scheduling and power gating for gpgpus,'' in \emph{Proceedings of the 46th Annual IEEE/ACM International Symposium on Microarchitecture (MICRO)}, 2013.

\bibitem{ORIGAMI}
M.~Abdel-Majeed, D.~Wong, J.~Kuang, and M.~Annavaram, ``Origami: Folding warps for energy efficient gpus,'' in \emph{Proceedings of the 30th ACM on International Conference on Supercomputing (ICS)}, 2016.

\bibitem{PRO}
J.~Anantpur and R.~Govindarajan, ``Pro: Progress aware gpu warp scheduling algorithm,'' in \emph{Proceedings of International Symposium on Parallel and Distributed Processing (IPDPS)}, 2015.

\bibitem{dwm}
E.~Atoofian, ``Reducing shift penalty in domain wall memory through register locality,'' in \emph{Proceedings of the International Conference on Compilers, Architectures and Synthesis for Embedded Systems (CASES)}, 2015.

\bibitem{STT-RAM-atof}
E.~Atoofian, ``A low power stt-ram based register file for gpgpus,'' in \emph{Proceedings of the 31st Annual ACM Symposium on Applied Computing (SAC)}, 2016.

\bibitem{PAWS}
M.~Awatramani, X.~Zhu, J.~Zambreno, and D.~Rover, ``Phase aware warp scheduling: Mitigating effects of phase behavior in gpgpu applications,'' in \emph{Proceedings of the International Conference on Parallel Architecture and Compilation Techniques (PACT)}, 2015.

\bibitem{Deepbench}
baidu, ``Deepbench: Benchmarking deep learning operations on different hardware,'' \url{https://github.com/baidu-research/DeepBench}, 2020.

\bibitem{CPU1}
R.~Balasubramonian, S.~Dwarkadas, and D.~H. Albonesi, ``Reducing the complexity of the register file in dynamic superscalar processors,'' in \emph{Proceedings of the 34th Annual IEEE/ACM International Symposium on Microarchitecture (MICRO)}, 2001.

\bibitem{SubCore}
A.~Barnes, F.~Shen, and T.~G. Rogers, ``Mitigating gpu core partitioning performance effects,'' in \emph{IEEE International Symposium on High-Performance Computer Architecture (HPCA)}, 2023.

\bibitem{CPU2}
E.~Borch, E.~Tune, S.~Manne, and J.~Emer, ``Loose loops sink chips,'' in \emph{Proceedings of the 8th IEEE International Symposium on High Performance Computer Architecture (HPCA)}, 2002.

\bibitem{Rodinia}
S.~Che, M.~Boyer, J.~Meng, D.~Tarjan, J.~W. Sheaffer, S.-H. Lee, and K.~Skadron, ``Rodinia: A benchmark suite for heterogeneous computing,'' in \emph{Proceedings of the IEEE International Symposium on Workload Characterization (IISWC)}, 2009.

\bibitem{GRGS}
J.~Chen, X.~Tao, Z.~Yang, J.-K. Peir, X.~Li, and S.-L. Lu, ``Guided region-based gpu scheduling: Utilizing multi-thread parallelism to hide memory latency,'' in \emph{Proceedings of the International Parallel and Distributed Processing Symposium (IPDPS)}, 2013.

\bibitem{CPU3}
J.-L. Cruz, A.~Gonzalez, M.~Valero, and N.~P. Topham, ``Multiple-banked register file architectures,'' in \emph{Proceedings of the 27th Annual International Symposium on Computer Architecture (ISCA)}, 2000.

\bibitem{STT-RAM-sched}
Q.~Deng, Y.~Zhang, Z.~Zhao, S.~Zhang, M.~Zhang, and J.~Yang, ``Frf: Toward warp-scheduler friendly stt-ram/sram fine-grained hybrid gpgpu register file design,'' \emph{IEEE Transactions on Computer-Aided Design of Integrated Circuits and Systems}, vol.~39, no.~10, pp. 2396--2409, 2020.

\bibitem{BOWS}
A.~ElTantawy and T.~M. Aamodt, ``Warp scheduling for fine-grained synchronization,'' in \emph{Proceedings of the 24th IEEE International Symposium on High Performance Computer Architecture (HPCA)}, 2018.

\bibitem{BOW}
H.~A. Esfeden, A.~Abdolrashidi, S.~Rahman, D.~Wong, and N.~Abu-Ghazaleh, ``Bow: Breathing operand windows to exploit bypassing in gpus,'' in \emph{Proceedings of the 53rd Annual IEEE/ACM International Symposium on Microarchitecture (MICRO)}, 2020.

\bibitem{Corf}
H.~A. Esfeden, F.~Khorasani, H.~Jeon, D.~Wong, and N.~Abu-Ghazaleh, ``Corf: Coalescing operand register file for gpus,'' in \emph{Proceedings of the 24th International Conference on Architectural Support for Programming Languages and Operating Systems (ASPLOS)}, 2019.

\bibitem{RFC}
M.~Gebhart, D.~R. Johnson, D.~Tarjan, S.~W. Keckler, W.~J. Dally, E.~Lindholm, and K.~Skadron, ``Energy-efficient mechanisms for managing thread context in throughput processors,'' in \emph{Proceedings of the 38th Annual International Symposium on Computer Architecture (ISCA)}, 2011.

\bibitem{RFC-Compiler}
M.~Gebhart, S.~W. Keckler, and W.~J. Dally, ``A compile-time managed multi-level register file hierarchy,'' in \emph{Proceedings of the 44th Annual IEEE/ACM International Symposium on Microarchitecture (MICRO)}, 2011.

\bibitem{RF-Virtualization}
H.~Jeon, G.~S. Ravi, N.~S. Kim, and M.~Annavaram, ``Gpu register file virtualization,'' in \emph{Proceedings of the 48th Annual IEEE/ACM International Symposium on Microarchitecture (MICRO)}, 2015.

\bibitem{DissectVolta}
Z.~Jia, M.~Maggioni, B.~Staiger, and D.~P. Scarpazza, ``Dissecting the nvidia volta gpu architecture via microbenchmarking,'' \url{https://arxiv.org/abs/1804.06826}, 2018.

\bibitem{edram}
N.~Jing, L.~Jiang, T.~Zhang, C.~Li, F.~Fan, and X.~Liang, ``Energy-efficient edram-based on-chip storage architecture for gpgpus,'' \emph{IEEE Transactions on Computers}, vol.~65, no.~1, pp. 122--135, 2016.

\bibitem{OSPG}
A.~Jog, O.~Kayiran, A.~K. Mishra, M.~T. Kandemir, O.~Mutlu, R.~Iyer, and C.~R. Das, ``Orchestrated scheduling and prefetching for gpgpus,'' in \emph{Proceedings of the 40th Annual International Symposium on Computer Architecture (ISCA)}, 2013.

\bibitem{OWL}
A.~Jog, O.~Kayiran, N.~C. Nachiappan, A.~K. Mishra, M.~T. Kandemir, O.~Mutlu, R.~Iyer, and C.~R. Das, ``Owl: cooperative thread array aware scheduling techniques for improving gpgpu performance,'' in \emph{Proceedings of the 18th International Conference on Architectural Support for Programming Languages and Operating Systems (ASPLOS)}, 2013.

\bibitem{accelwattch}
V.~Kandiah, S.~Peverelle, M.~Khairy, J.~Pan, A.~Manjunath, T.~G. Rogers, T.~M. Aamodt, and N.~Hardavellas, ``Accelwattch: A power modeling framework for modern gpus,'' in \emph{Proceedings of the 54th Annual IEEE/ACM International Symposium on Microarchitecture (MICRO)}, 2021.

\bibitem{Accelsim}
M.~Khairy, Z.~Shen, T.~M. Aamodt, and T.~G. Rogers, ``Accel-sim: An extensible simulation framework for validated gpu modeling,'' in \emph{Proceedings of the 47th Annual International Symposium on Computer Architecture (ISCA)}, 2020.

\bibitem{Regless}
J.~Kloosterman, J.~Beaumont, D.~A. Jamshidi, J.~Bailey, T.~Mudge, and S.~Mahlke, ``Regless: Just-in-time operand staging for gpus,'' in \emph{Proceedings of the 50th Annual IEEE/ACM International Symposium on Microarchitecture (MICRO)}, 2017.

\bibitem{CAPAS}
G.~Koo, H.~Jeon, Z.~Liu, N.~S. Kim, and M.~Annavaram, ``Cta-aware prefetching and scheduling for gpu,'' in \emph{Proceedings of the International Parallel and Distributed Processing Symposium (IPDPS)}, 2018.

\bibitem{IPAWS}
M.~Lee, G.~Kim, J.~Kim, W.~Seo, Y.~Cho, and S.~Ryu, ``ipaws : Instruction-issue pattern-based adaptive warp scheduling for gpgpus,'' in \emph{Proceedings of the 22nd IEEE International Symposium on High Performance Computer Architecture (HPCA}, 2016.

\bibitem{IGRUT}
M.~Lee, S.~Song, J.~Moon, J.~Kim, W.~Seo, Y.~Cho, and S.~Ryu, ``Improving gpgpu resource utilization through alternative thread block scheduling,'' in \emph{Proceedings of the 20th IEEE International Symposium on High Performance Computer Architecture (HPCA)}, 2014.

\bibitem{Warped-Compression}
S.~Lee, K.~Kim, G.~Koo, H.~Jeon, W.~W. Ro, and M.~Annavaram, ``Warped-compression: Enabling power efficient gpus through register compression,'' in \emph{Proceedings of the 42nd Annual International Symposium on Computer Architecture (ISCA)}, 2015.

\bibitem{CAWA}
S.-Y. Lee, A.~Arunkumar, and C.-J. Wu, ``Cawa: Coordinated warp scheduling and cache prioritization for critical warp acceleration of gpgpu workloads,'' in \emph{Proceedings of the 42nd Annual International Symposium on Computer Architecture (ISCA)}, 2015.

\bibitem{CAWS}
S.-Y. Lee and C.-J. Wu, ``Caws: Criticality-aware warp scheduling for gpgpu workloads,'' in \emph{Proceedings of the 23rd International Conference on Parallel Architecture and Compilation Techniques (PACT)}, 2014.

\bibitem{STT-RAM-Gushu}
G.~Li, X.~Chen, G.~Sun, H.~Hoffmann, Y.~Liu, Y.~Wang, and H.~Yang, ``A stt-ram-based low-power hybrid register file for gpgpus,'' in \emph{Proceedings of the 52nd ACM/IEEE Design Automation Conference (DAC)}, 2015.

\bibitem{Tesla}
E.~Lindholm, J.~Nickolls, S.~Oberman, and J.~Montrym, ``Nvidia tesla: A unified graphics and computing architecture,'' \emph{IEEE Micro}, vol.~28, no.~2, pp. 39--55, 2008.

\bibitem{SAWS}
J.~Liu, J.~Yang, and R.~Melhem, ``Saws: Synchronization aware gpgpu warp scheduling for multiple independent warp schedulers,'' in \emph{Proceedings of the 48th Annual IEEE/ACM International Symposium on Microarchitecture (MICRO)}, 2015.

\bibitem{OCU}
S.~Liu, J.~E. Lindholm, M.~Y. Siu, B.~W. Coon, and S.~F. Oberman, ``Operand collector architecture,'' U.S. Patent 7834881 B2, Nov. 16, 2010.

\bibitem{BAWS}
Y.~Liu, Z.~Yu, L.~Eeckhout, V.~J. Reddi, Y.~Luo, X.~Wang, Z.~Wang, and C.~Xu, ``Barrier-aware warp scheduling for throughput processors,'' in \emph{Proceedings of the 30th ACM on International Conference on Supercomputing (ICS)}, 2016.

\bibitem{VWS}
M.~Mao, J.~Hu, Y.~Chen, and H.~Li, ``Vws: A versatile warp scheduler for exploring diverse cache localities of gpgpu applications,'' in \emph{Proceedings of the 52nd ACM/IEEE Design Automation Conference (DAC)}, 2015.

\bibitem{RT1}
M.~Mao, W.~Wen, Y.~Zhang, Y.~Chen, and H.~Li, ``Exploration of gpgpu register file architecture using domain-wall-shift-write based racetrack memory,'' in \emph{Proceedings of the 51st ACM/IEEE Design Automation Conference (DAC)}, 2014.

\bibitem{RT}
M.~Mao, W.~Wen, Y.~Zhang, Y.~Chen, and H.~Li, ``An energy-efficient gpgpu register file architecture using racetrack memory,'' \emph{IEEE Transactions on Computers}, vol.~66, no.~9, pp. 1478--1490, 2017.

\bibitem{LargeWarp}
V.~Narasiman, M.~Shebanow, C.~J. Lee, R.~Miftakhutdinov, O.~Mutlu, and Y.~N. Patt, ``Improving gpu performance via large warps and two-level warp scheduling,'' in \emph{Proceedings of the 44th Annual IEEE/ACM International Symposium on Microarchitecture (MICRO)}, 2011.

\bibitem{CPU4}
P.~R. Nuth and W.~J. Dally, ``The named-state register file: implementation and performance,'' in \emph{Proceedings of the 1st IEEE International Symposium on High Performance Computer Architecture (HPCA)}, 1995.

\bibitem{SASS}
NVIDIA, ``cuda-binary-utilities release 12.1,'' \url{https://docs.nvidia.com/cuda/pdf/CUDA_Binary_Utilities.pdf}.

\bibitem{CudaGuide}
NVIDIA, ``Cuda c++ programming guide,'' \url{https://docs.nvidia.com/cuda/pdf/CUDA_C_Programming_Guide.pdf}.

\bibitem{A100whitepaper}
NVIDIA, ``Nvidia a100 tensor core gpu architecture,'' \url{https://images.nvidia.com/aem-dam/en-zz/Solutions/data-center/nvidia-ampere-architecture-whitepaper.pdf}.

\bibitem{Adawhitepaper}
NVIDIA, ``Nvidia ada gpu architecture,'' \url{https://images.nvidia.com/aem-dam/Solutions/geforce/ada/nvidia-ada-gpu-architecture.pdf}.

\bibitem{GTX750whitepaper}
NVIDIA, ``Nvidia geforce gtx 750 ti,'' \url{https://international.download.nvidia.com/geforce-com/international/pdfs/GeForce-GTX-750-Ti-Whitepaper.pdf}.

\bibitem{Maxwellwhitepaper}
NVIDIA, ``Nvidia geforce gtx 980,'' \url{https://www.microway.com/download/whitepaper/NVIDIA_Maxwell_GM204_Architecture_Whitepaper.pdf}.

\bibitem{H100whitepaper}
NVIDIA, ``Nvidia h100 tensor core gpu architecture,'' \url{https://resources.nvidia.com/en-us-tensor-core/gtc22-whitepaper-hopper}.

\bibitem{P100whitepaper}
NVIDIA, ``Nvidia tesla p100,'' \url{https://images.nvidia.com/content/pdf/tesla/whitepaper/pascal-architecture-whitepaper.pdf}.

\bibitem{V100whitepaper}
NVIDIA, ``Nvidia tesla v100 gpu architecture,'' \url{https://images.nvidia.com/content/volta-architecture/pdf/volta-architecture-whitepaper.pdf}.

\bibitem{Turingwhitepaper}
NVIDIA, ``Nvidia turing gpu architecture,'' \url{https://images.nvidia.com/aem-dam/en-zz/Solutions/design-visualization/technologies/turing-architecture/NVIDIA-Turing-Architecture-Whitepaper.pdf}.

\bibitem{TU100whitepaper}
NVIDIA, ``Nvidia turing gpu architecture,'' \url{https://images.nvidia.com/aem-dam/en-zz/Solutions/design-visualization/technologies/turing-architecture/NVIDIA-Turing-Architecture-Whitepaper.pdf}.

\bibitem{PTX}
NVIDIA, ``Ptx isa release 8.1,'' \url{https://docs.nvidia.com/cuda/pdf/ptx_isa_8.1.pdf}.

\bibitem{CUDA}
NVIDIA, ``Cuda c++ programming guide,'' Technical Report PG-02829-001\_v11.7, 2022.

\bibitem{ELF}
J.~J.~K. Park, Y.~Park, and S.~Mahlke, ``Elf: maximizing memory-level parallelism for gpus with coordinated warp and fetch scheduling,'' in \emph{Proceedings of the International Conference for High Performance Computing, Networking, Storage and Analysis (SC)}, 2015.

\bibitem{TENSOR}
M.~A. Raihan, N.~Goli, and T.~M. Aamodt, ``Modeling deep learning accelerator enabled gpus,'' in \emph{Proceedings of the IEEE International Symposium on Performance Analysis of Systems and Software (ISPASS)}, 2019.

\bibitem{CCWS}
T.~G. Rogers, M.~O'Connor, and T.~M. Aamodt, ``Cache-conscious wavefront scheduling,'' in \emph{Proceedings of the 45th Annual IEEE/ACM International Symposium on Microarchitecture (MICRO)}, 2012.

\bibitem{DAWS}
T.~G. Rogers, M.~O'Connor, and T.~M. Aamodt, ``Divergence-aware warp scheduling,'' in \emph{Proceedings of the 46th Annual IEEE/ACM International Symposium on Microarchitecture (MICRO)}, 2013.

\bibitem{LTRF}
M.~Sadrosadati, A.~Mirhosseini, S.~B. Ehsani, H.~Sarbazi-Azad, M.~Drumond, B.~Falsafi, R.~Ausavarungnirun, and O.~Mutlu, ``Ltrf: Enabling high-capacity register files for gpus via hardware/software cooperative register prefetching,'' in \emph{Proceedings of the 23rd International Conference on Architectural Support for Programming Languages and Operating Systems (ASPLOS)}, 2018.

\bibitem{CPU5}
R.~Shioya, K.~Horio, M.~Goshima, and S.~Sakai, ``Register cache system not for latency reduction purpose,'' in \emph{Proceedings of the 43rd Annual IEEE/ACM International Symposium on Microarchitecture}, 2010.

\bibitem{QAWS}
J.~Singh, I.~S. Olmedo, N.~Capodieci, A.~Marongiu, and M.~Caccamo, ``Reconciling qos and concurrency in nvidia gpus via warp-level scheduling,'' in \emph{Proceedings of the Design, Automation \& Test in Europe Conference \& Exhibition (DATE)}, 2022.

\bibitem{Reliabilty-Combating}
J.~Tan, S.~L. Song, K.~Yan, X.~Fu, A.~Marquez, and D.~Kerbyson, ``Combating the reliability challenge of gpu register file at low supply voltage,'' in \emph{Proceedings of the International Conference on Parallel Architecture and Compilation Techniques (PACT)}, 2016.

\bibitem{OAWS}
B.~Wang, Y.~Zhu, and W.~Yu, ``Oaws: Memory occlusion aware warp scheduling,'' in \emph{Proceedings of the International Conference on Parallel Architecture and Compilation Techniques (PACT)}, 2016.

\bibitem{STT-RAM1}
J.~Wang and Y.~Xie, ``A write-aware sttram-based register file architecture for gpgpu,'' \emph{J. Emerg. Technol. Comput. Syst.}, vol.~12, no.~1, aug 2015.

\bibitem{PATS}
Q.~Xu and M.~Annavaram, ``Pats: Pattern aware scheduling and power gating for gpgpus,'' in \emph{Proceedings of the 23rd International Conference on Parallel Architecture and Compilation Techniques (PACT)}, 2014.

\bibitem{TENSOR1}
D.~Yan, W.~Wang, and X.~Chu, ``Demystifying tensor cores to optimize half-precision matrix multiply,'' in \emph{Proceedings of the IEEE International Parallel and Distributed Processing Symposium (IPDPS))}, 2020.

\bibitem{SAW}
Y.~Yu, W.~Xiao, X.~He, H.~Guo, Y.~Wang, and X.~Chen, ``A stall-aware warp scheduling for dynamically optimizing thread-level parallelism in gpgpus,'' in \emph{Proceedings of the 29th ACM on International Conference on Supercomputing (ICS)}, 2015.

\bibitem{SinglePortedRF}
L.~Yue, J.~W. Berendsen, K.~M. Abdalla, R.~M. Bastos, and R.~Danilak, ``Architecture for compact multi-ported register file,'' U.S. Patent 7490208 B1, Feb. 10, 2009.

\bibitem{CPU6}
H.~Zeng and K.~Ghose, ``Register file caching for energy efficiency,'' in \emph{Proceedings of the International Symposium on Low Power Electronics and Design (ISLPED)}, 2006.

\bibitem{CIAO}
J.~Zhang, S.~Gao, N.~S. Kim, and M.~Jung, ``Ciao: Cache interference-aware throughput-oriented architecture and scheduling for gpus,'' in \emph{Proceedings of the International Parallel and Distributed Processing Symposium (IPDPS)}, 2018.

\end{thebibliography}

\end{document}